\documentclass[aps,prl,longbibliography,showpacs,twocolumn,superscriptaddress]{revtex4-2}
\usepackage{amsmath,amssymb,amsfonts,bm}
\usepackage{booktabs}
\usepackage{braket}
\usepackage{graphicx}
\usepackage{epstopdf}
\usepackage{dcolumn}
\usepackage{mathrsfs}
\usepackage{bbold}
\usepackage{dsfont}
\usepackage{float}

\usepackage{tgtermes}
\usepackage{multirow}
\usepackage[colorlinks=true,linkcolor=blue,citecolor=blue, urlcolor=blue,bookmarks=false]{hyperref}

\begin{document}
	
\title{Floquet second-order topological insulator in strained graphene}
\author{Yu-Wen Xu}
\thanks{These authors contributed equally to this work.}
\affiliation{Department of Physics and Chongqing Key Laboratory for Strongly Coupled Physics, Chongqing University, Chongqing 400044,  China}

\author{Xiaolin Wan}
\thanks{These authors contributed equally to this work.}
\affiliation{Department of Physics and Chongqing Key Laboratory for Strongly Coupled Physics, Chongqing University, Chongqing 400044,  China}

\author{Zi-Ming Wang}
\thanks{These authors contributed equally to this work.}
\affiliation{Department of Physics and Chongqing Key Laboratory for Strongly Coupled Physics, Chongqing University, Chongqing 400044,  China}

\author{Rui Wang}
\affiliation{Department of Physics and Chongqing Key Laboratory for Strongly Coupled Physics, Chongqing University, Chongqing 400044, China}
\affiliation{Center of Quantum Materials and Devices, Chongqing University, Chongqing 400044, China}

\author{Dong-Hui Xu}
\email{donghuixu@cqu.edu.cn}
\affiliation{Department of Physics and Chongqing Key Laboratory for Strongly Coupled Physics, Chongqing University, Chongqing 400044, China}
\affiliation{Center of Quantum Materials and Devices, Chongqing University, Chongqing 400044, China}


\begin{abstract}
Graphene provides a canonical setting for Floquet band engineering, where circularly polarized light can dynamically open topological gaps at Dirac points and generate nonequilibrium Hall responses. Here we show that uniaxial strain and off-resonant circularly polarized light with tunable incidence angle enable a controllable route to Floquet higher-order topology in graphene. Using a strained honeycomb tight-binding model with Peierls coupling and a high-frequency expansion for the effective Floquet Hamiltonian, we find that strain drives the Dirac cones toward the Dirac-merging (semi-Dirac) critical regime, where the light-induced mass becomes strongly anisotropic. For oblique incidence, the projected drive is effectively elliptically polarized and, in combination with strain, stabilizes a phase with gapped edges but robust in-gap corner modes in finite geometries, realizing a Floquet second-order topological insulator. We characterize the phase diagram via the Chern number and a crystalline-symmetry-quantized polarization invariant. Finally, first-principles-informed tight-binding calculations corroborate the predicted topological evolution in strained graphene nanostructures. Our results identify driven strained graphene as a realistic and tunable platform for realizing and diagnosing Floquet higher-order topological phases.

\end{abstract}
\maketitle
	
\emph{Introduction.}---Two-dimensional ($2\text{D}$) Dirac semimetals (DSMs) serve as a foundational platform for topological quantum phenomena. These materials host low-energy quasiparticles that behave as massless Dirac fermions, exhibiting characteristic responses driven by the Berry phase \cite{Dirac-review}. Graphene \cite{Graphene-Review}, the prototypical $2\text{D}$ DSM, features Dirac cones at the Brillouin zone (BZ) corners and provided the initial framework for the theoretical discovery of the quantum anomalous Hall (QAH) \cite{Haldane1988} and quantum spin Hall (QSH) \cite{KMmodel} effects. In these models, the transition to a topological phase is achieved by generating a Dirac mass term (staggered flux or intrinsic spin–orbit coupling), highlighting how symmetry constrains permissible masses and thereby dictates band topology and quantized boundary responses.

A particularly powerful extension of this paradigm is Floquet engineering, where periodic driving reshapes electronic bands and generates effective Hamiltonians with drive-controlled topology~\cite{pssr.201206451,annurev-conmatphys-031218-013423,harper-annurev-conmatphys,rudner2020band,bao2022light,PhysRevB.79.081406,PhysRevLett.105.017401,PhysRevB.82.235114,PhysRevLett.107.216601,PhysRevB.84.235108,PhysRevB.88.245422,PhysRevLett.113.266801,PhysRevB.90.115423,lindner2011floquet,PhysRevA.82.033429,PhysRevB.87.235131,PhysRevLett.106.220402,PhysRevX.3.031005,PhysRevLett.110.026603,PhysRevLett.110.200403,PhysRevLett.112.156801,Wang_2014,PhysRevB.94.041409,PhysRevLett.117.087402,PhysRevB.94.155206,PhysRevLett.116.156803,PhysRevB.91.205445,PhysRevB.94.081103,PhysRevB.94.121106,PhysRevB.96.041206,PhysRevB.93.205435,hubener2017creating,PhysRevB.97.155152,PhysRevB.97.195139,PhysRevB.100.165302,PhysRevB.102.201105,Ghosh2022systematic,nag2021anomalous,Nag2019dynamical,XLDU2022,ZMWangFloquet,zhan2024perspective,Zengprl,PhysRevLett.128.066602,PhysRevB.99.045441,PhysRevB.100.085138,PhysRevResearch.1.032045,PhysRevB.100.115403,PhysRevResearch.1.032045,PhysRevLett.123.016806,PhysRevLett.124.057001,PhysRevB.101.235403,PhysRevResearch.2.013124,PhysRevResearch.2.033495,PhysRevLett.124.216601,PhysRevB.103.L041402,PhysRevB.103.L121115,PhysRevB.103.115308,PhysRevB.104.L020302,PhysRevResearch.3.L032026,PhysRevB.103.184510,PhysRevB.104.205117,Ghosh2022hinge-mode,Liupprb,wanprb1,Huangprb}. In graphene, circularly polarized light (CPL) provides a conceptually clean route: it dynamically breaks time-reversal symmetry and generates an effective Dirac mass, opening gaps at the Dirac points and enabling Floquet Chern phases with chiral edge states~\cite{PhysRevB.79.081406,PhysRevB.84.235108,annurev-conmatphys-031218-013423}. Experimentally, progress has advanced from transport signatures to energy–momentum-resolved spectroscopy: ultrafast on-chip measurements reported a helicity-dependent anomalous Hall response in graphene under femtosecond CPL with gate dependence consistent with light-induced band restructuring~\cite{mciver2020light}, while recent time-resolved photoemission directly observed light-dressed Dirac dispersions and Floquet–Bloch replica features and developed analyses to disentangle initial-state Floquet dressing from photon-dressed final-state (Volkov) effects~\cite{Choi2025,Merboldt2025}. More recently, a Floquet-induced gap in graphene was directly observed by time-resolved photoemission~\cite{Wang2026}. These advances motivate the pursuit of nonequilibrium topological phases in graphene beyond first-order Chern insulators.

Despite this progress, an important frontier remains largely unexplored: Floquet higher-order topology in graphene. As the classification of topological matter has extended beyond the conventional bulk–boundary correspondence, increasing attention has been devoted to higher-order topological insulators (HOTIs) \cite{Zhang2013PRL,Benalcazar2017Science,Langbehn2017PRL,Song2017PRL,Schindler2018SA,PhysRevLett.124.036803,xie2021higher}, which host boundary states of co-dimension two—0D corner modes rather than conventional 1D edge channels. Realizing such phases in a Floquet setting is qualitatively more demanding than generating first-order phases, as it requires opening a bulk gap while simultaneously producing a symmetry-constrained boundary mass structure that enforces sign changes of the boundary gap on adjacent edges, thereby pinning corner-localized states.

In this work, we propose an optically driven route to realize and switch a Floquet second-order topological insulating (SOTI) phase in graphene by combining:
(i) uniaxial strain, which reshapes and displaces the Dirac cones and enables access to the Dirac-merging (semi-Dirac) regime; and (ii) off-resonant CPL with a tunable incidence direction, which controls the effective in-plane driving geometry via projection. We show that strained graphene under oblique-incidence CPL can undergo a transition from a conventional Floquet Chern phase into a Floquet SOTI characterized by gapped edges but robust in-gap corner modes in finite geometries. The semi-Dirac regime renders the drive-induced mass strongly anisotropic and orientation dependent, promoting an edge-dependent boundary-mass pattern and a Floquet mass-domain-wall mechanism at corners. We map the phase diagram using the Chern number together with a symmetry-quantized polarization invariant protected by an antiunitary crystalline symmetry that survives strain and oblique driving. Finally, first-principles-informed Wannier tight-binding simulations corroborate the predicted phase sequence in strained graphene nanostructures.

\begin{figure}[t!]
    \centering
    \includegraphics[width=0.9\linewidth]{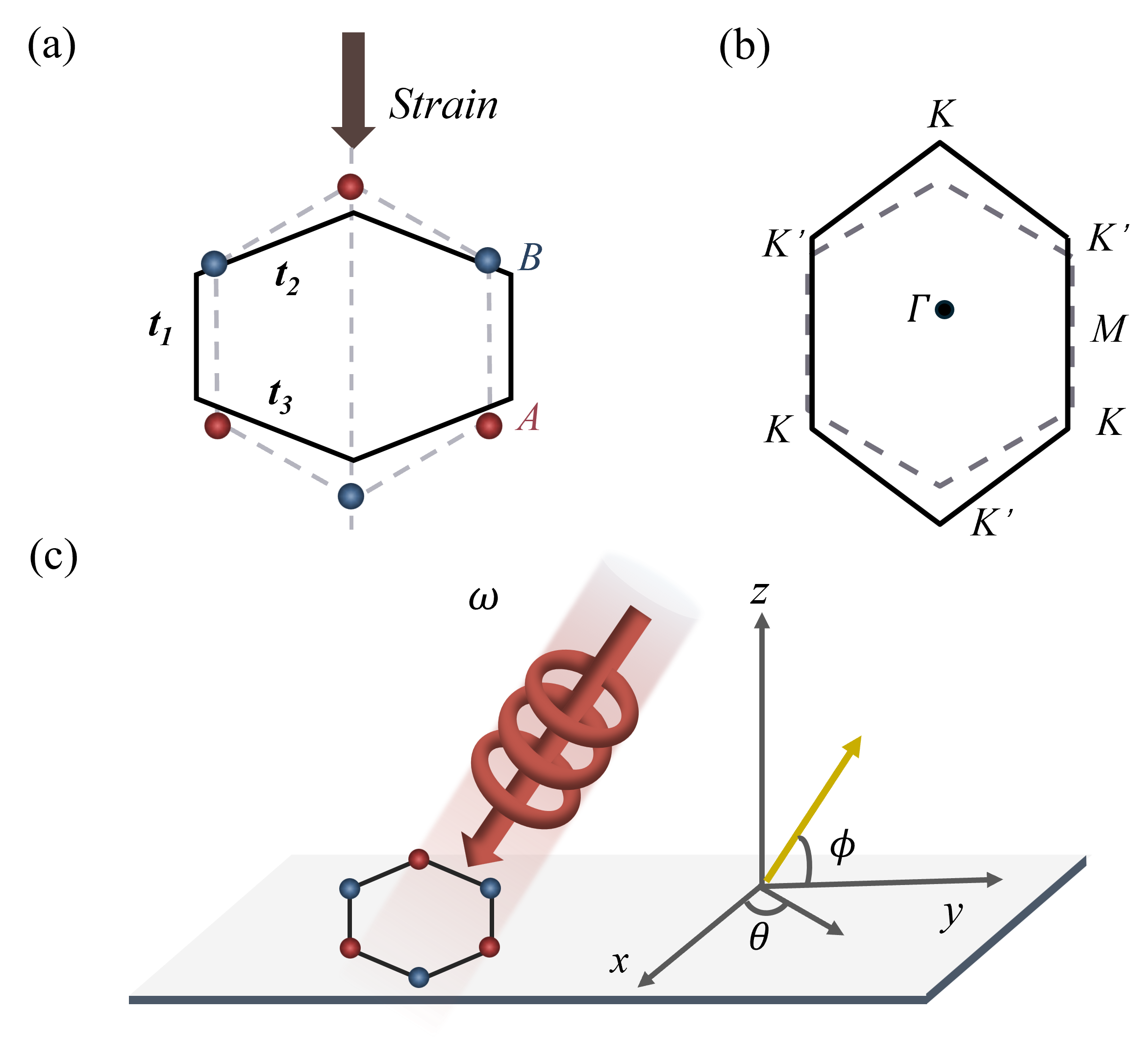}
    \caption{Uniaxial strain and driven-graphene geometry. (a) Uniaxially strained honeycomb lattice: strain is applied along $y$, generating anisotropic hoppings ($t_1\neq t_2=t_3$). (b) Schematic BZ deformation under strain and the associated shift of the Dirac nodes along the strain axis. (c) Circularly polarized light with tunable propagation direction parameterized by polar angle $\phi$ (the elevation angle measured from the graphene plane) and azimuthal angle $\theta$. For oblique incidence ($\phi\neq\pi/2$), the in-plane projection of the drive is effectively elliptically polarized.}
    \label{fig1}
\end{figure}

\textit{Model and Floquet theory.}---We model unstrained graphene by a spinless nearest-neighbor tight-binding Hamiltonian on the honeycomb lattice
\begin{equation}
H=-\sum_{\langle ij \rangle}\left[ t_{ij} c_{i}^\dagger c_{j} + \text{h.c.} \right],
\end{equation}
where $c_i^\dagger$ creates an electron on site $i$ and $\langle ij\rangle$ runs over nearest neighbors. Uniaxial strain is encoded through strain-dependent hoppings: $t_{ij}=t_{\text{0}} \exp \left[ {-\beta\left ( \tfrac{d_{ij}}{d_{\text{0}}}-1 \right ) } \right]$ with $t_{0}$ and $d_{0}$ the unstrained hopping and bond length, respectively, and $\beta$ the Gr\"uneisen parameter. We take the three nearest-neighbor vectors as $\bm a_1=(0,-1)$, $\bm a_2=(\sqrt{3}/2,1/2)$, and $\bm a_3=(-\sqrt{3}/2,1/2)$ (bond length set to unity). For a uniaxial strain $\delta$ applied along $y$ [compression along $y$ and Poisson expansion along $x$, Fig.~\ref{fig1}(a)], the strained bond length reads
$d_{ij}=\sqrt{(1+\nu\delta)^2(\bm r_{ij}\!\cdot\!\bm e_x)^2+(1-\delta)^2(\bm r_{ij}\!\cdot\!\bm e_y)^2}$, where $\bm r_{ij}=\bm r_i-\bm r_j$, $\{\bm e_x,\bm e_y\}$ are unit vectors, and we use $\nu=0.165$ for graphene~\cite{C5NR07755A,PhysRevB.81.081407,Naumis_2017,Pereira_2010,PhysRevB.78.075435}. Within the experimentally accessible strain range of graphene, typically about $15\%$ to above $20\%$~\cite{Lee2008Sci,Wei2009PRB}, we adopt a representative effective strain of $\epsilon \approx 10\%$. In
the following, we use the natural units $e=\hbar=c=1$.

Unstrained graphene (identical sublattices) has point group $D_6$; uniaxial strain lowers it to $D_2$ generated by $\{{ C}_{2x}, {C}_{2z}\}$. In particular, removing ${ C}_{3z}$ unpins the Dirac nodes from the BZ corners and allows them to move and eventually merge along the strain axis, approaching the semi-Dirac~\cite{SemiDirac1,SemiDirac2} regime.

\begin{figure}[t!]
    \centering
    \includegraphics[width=1\linewidth]{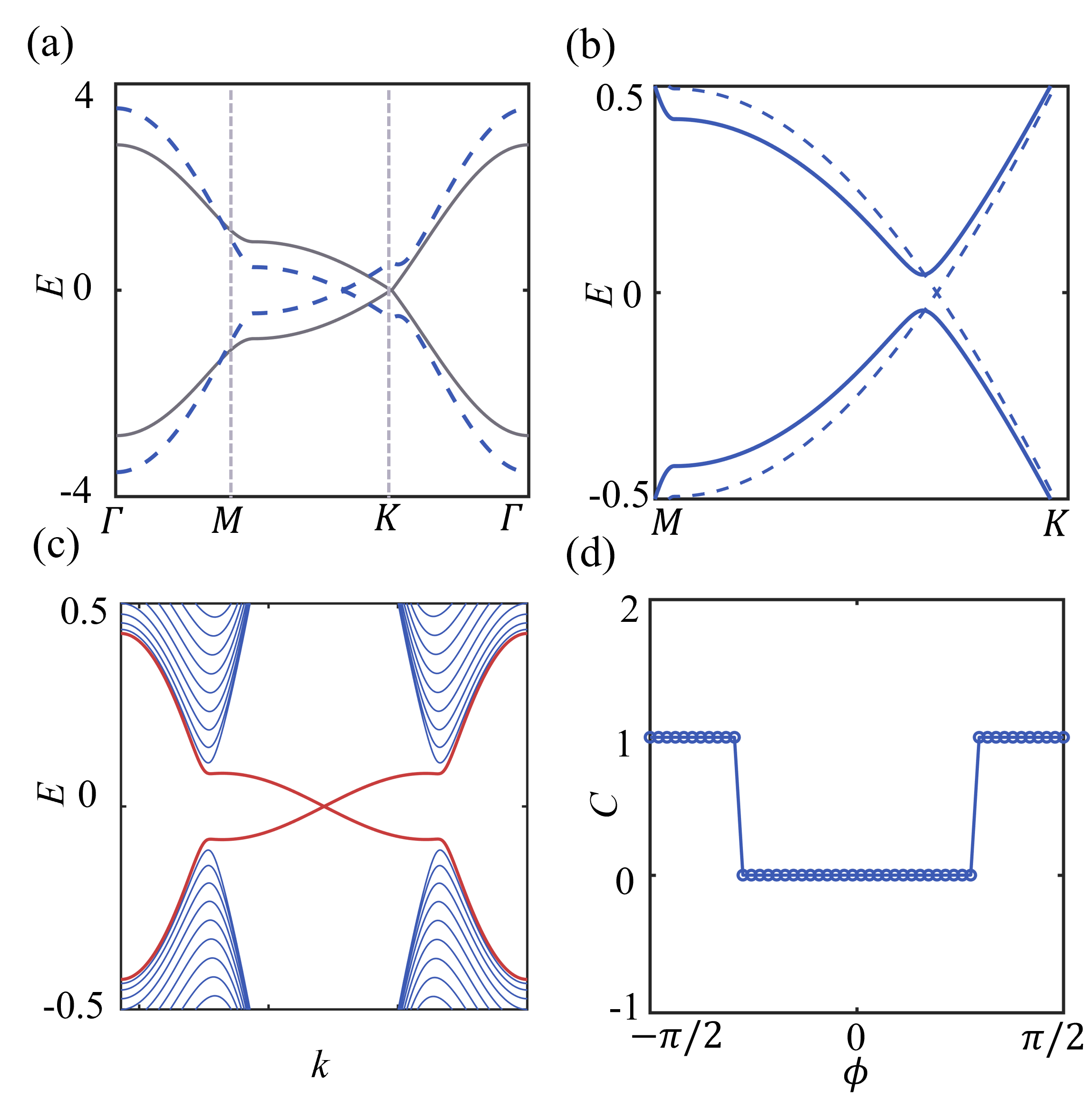}
    \caption{Floquet gap opening and first-order topology.
(a) Bulk bands of strained graphene along $\Gamma$--$M$--$K$--$\Gamma$
(unstrained: gray; strained: dashed blue). Parameters: $\delta=0.1$, and $\beta=4.8$. (b) Zoom-in showing a light-induced gap at the strain-shifted Dirac points. Parameters: $t_0=1,\omega=3,A_0=0.5,\phi=\frac{\pi}{3}$ and $\theta =\frac{\pi}{2}$. (c) Zigzag-ribbon quasienergy spectrum exhibiting a single chiral edge mode traversing the Floquet gap in the Chern phase. (d) Chern number $C$ versus incidence angle $\phi$: quantized plateaus indicate distinct Floquet phases; jumps coincide with gap closings.}
    \label{fig2}
\end{figure}

To describe irradiation by CPL with a tunable incidence direction, we introduce a time-periodic vector potential $\bm A(\tau)=\bm A(\tau+T)$ with $T=2\pi/\omega$, parameterized by the azimuthal and polar angles $(\theta,\phi)$ of the propagation direction [Fig.~\ref{fig1}(c)]:
\begin{align*}
A_x(\tau,\phi,\theta)&=A_0\cos\theta\,\sin(\omega\tau)-\eta A_0\sin\phi\,\sin\theta\,\cos(\omega\tau),\nonumber\\
A_y(\tau,\phi,\theta)&=A_0\sin\theta\,\sin(\omega\tau)+\eta A_0\sin\phi\,\cos\theta\,\cos(\omega\tau),\nonumber\\
A_z(\tau,\phi)&=\eta A_0\cos\phi\,\cos(\omega\tau),
\end{align*}
where $A_0$ is the field amplitude and $\eta=\pm1$ labels the helicity. The polar angle $\phi$ interpolates from grazing incidence ($\phi=0$) to normal incidence ($\phi=\pi/2$). As useful limiting cases, the above parameterization reduces to propagation along $x$ for $\phi=0,\theta=\pi/2$, $\bm A(\tau)=(0,\;A_0\sin\omega\tau,\;\eta A_0\cos\omega\tau)$; along $y$ for $\phi=0,\theta=\pi$, $\bm A(\tau)=(-A_0\sin\omega\tau,\;0,\;\eta A_0\cos\omega\tau)$; and along $z$ for $\phi=\pi/2,\theta=0$, $\bm A(\tau)=(A_0\sin\omega\tau,\;\eta A_0\cos\omega\tau,\;0)$. For 2D graphene, the Peierls phases on in-plane bonds depend only on the in-plane component; consequently, normal incidence ($\phi=\pi/2$) yields a purely in-plane circular drive, whereas oblique incidence ($\phi\neq\pi/2$) generally produces an elliptically polarized in-plane projection, which is essential for generating an edge-dependent mass pattern in the strained semi-Dirac regime.

Under this drive, the hopping amplitudes acquire Peierls phases, $t_{ij} \rightarrow t_{ij} \text{exp} \left[ -i \int_{\bm{r}_i}^{\bm{r}_j} \bm{A}(\tau) \cdot d\bm{r} \right]$, so that the Hamiltonian becomes periodic in time ${ H}(\tau)={ H}(\tau + T)$. Within the Floquet formalism~\cite{PhysRev.138.B979,PhysRevA.7.2203}, we introduce the Fourier components ${H}_l=\tfrac{1}{T}\int_{0}^{T}e^{il \omega \tau} { H}(\tau),\; l\in\mathbb{Z}$. We focus on the high-frequency (off-resonant) regime, where the photon energy is large enough that resonant interband transitions are strongly suppressed. In this limit, the effective static Floquet Hamiltonian follows from the high-frequency (van Vleck) expansion~\cite{bukov2015universal,Eckardt_2015}
\begin{equation}
{\cal H}_{\text{eff}}={\cal H}_{0}+\sum_{l\neq0}\frac{\big[{\cal H}_{-l},{\cal H}_{l}\big]}{l\,\omega}+{\cal O}(\omega^{-2}),
\label{eq:HFexp}
\end{equation}
and we retain the leading $l=\pm1$ contribution. The resulting effective Floquet Hamiltonian in reciprocal space takes the two-band form
\begin{align}\label{eq:Heff}
   \mathcal{H}_{\text{eff}} = J_0\sigma_1 M(\bm{k}) + J_0\sigma_2 G(\bm{k}) + \frac{\eta J_1^2 \sin\phi}{\omega}\sigma_3 W(\bm{k}), 
\end{align}
where $\sigma_{j=x,y,z}$ are the Pauli matrices acting on the sublattice degree of freedom, and $J_q(A_0)$ denotes the $q$-th order Bessel function of the first kind. The structure factors are $M(\bm{k}) = \sum_{i=1}^3 t_i \cos(\bm{k} \cdot \bm{a}_i)$, $G(\bm{k}) = \sum_{i=1}^3 t_i \sin(\bm{k} \cdot \bm{a}_i)$, and $W(\bm{k}) = -t_1 t_2 \sin(\bm{k} \cdot \bm{b}_1) + t_1 t_3 \sin(\bm{k} \cdot \bm{b}_2) + t_2 t_3 \sin(\bm{k} \cdot (\bm{b}_1 - \bm{b}_2))$. Here, $\bm{b}_1=(\frac{\sqrt{3}}{2},\frac{3}{2})$, 
$\bm{b}_2=(-\frac{\sqrt{3}}{2},\frac{3}{2})$, 
and $\bm{b}_3=(\sqrt{3},0)$ represent the next-nearest-neighbor vectors.  The last term in Eq.~\eqref{eq:Heff} is a Haldane-type Floquet mass that breaks time-reversal symmetry and scales as $\propto \sin\phi$; consequently, normal incidence maximizes the Chern gap, whereas grazing incidence suppresses it within the high-frequency description.

\textit{Floquet Chern insulator.}---We first consider CPL incident close to the surface normal, i.e., propagation along the $z$ direction ($\phi=\pi/2$), for which the in-plane component $\bm A_{\parallel}(\tau)=(A_x,A_y)$ is purely circular. The drive breaks time-reversal symmetry $\mathcal{T}$ and generates a Haldane-type mass term in ${\cal H}_{\rm eff}$, so that the Dirac crossings are generically gapped once the Floquet mass is nonzero. From the symmetry viewpoint, normal-incidence CPL breaks the in-plane twofold rotations ${ C}_{2x}$ and ${C}_{2y}$, reducing the point group from $D_6$ to $C_6$ in unstrained graphene and from $D_2$ to $C_2$ in the strained lattice; thus no crystalline symmetry remains that can protect the Dirac nodes against gapping. 

Strain modifies the location and anisotropy of the low-energy Dirac cones without changing this basic mechanism. In the unstrained limit the nearest-neighbor hoppings are uniform and the Dirac points reside at the BZ corners $K$ and $K'$. Under uniaxial strain the BZ is deformed and the Dirac points shift away from $K/K'$ along the strain axis, as shown in Fig.~\ref{fig2}(a). Upon irradiation, a Floquet gap opens at the (shifted) Dirac points [Fig.~\ref{fig2}(b)], and the corresponding ribbon spectrum exhibits a single chiral edge mode traversing the gap [Fig.~\ref{fig2}(c)], characteristic of a Floquet Chern insulator. We quantify the first-order topology by the Chern number of the occupied Floquet band
\begin{equation}
\mathcal{C}=\frac{1}{2\pi}\int_{\rm BZ} d^2k\,\Omega_z(\bm k),
\end{equation}
where $\Omega_z(\bm k)$ is the Berry curvature. Figure~\ref{fig2}(d) shows $\mathcal{C}$ versus the incidence angle $\phi$. In the grazing-incidence limit ($|\phi|\ll1$), the Floquet mass is strongly suppressed (in our high-frequency expansion it scales as $\propto\sin\phi$), and the system remains topologically trivial ($\mathcal{C}=0$). As $\phi$ approaches normal incidence ($\phi\to\pi/2$), the mass increases and a robust plateau with $\mathcal{C}=1$ develops, with topological transitions occurring only when the quasienergy gap closes.

Oblique incidence ($\phi\neq\pi/2$) changes the effective in-plane drive from circular to elliptic: although the electromagnetic field is circular in the plane perpendicular to the propagation direction, its projection onto the graphene plane is elliptically polarized. This projected drive breaks the threefold rotation ${C}_{3z}$ while preserving ${C}_{2z}$, so that both strained and unstrained honeycomb lattices reduce to a $C_2$ point group. As we show next, in the strained system near the Dirac-merging regime this reduced symmetry gives rise to orientation-dependent boundary masses, which are the key ingredient for higher-order topology.

\begin{figure}[t!]
    \centering
    \includegraphics[width=1\linewidth]{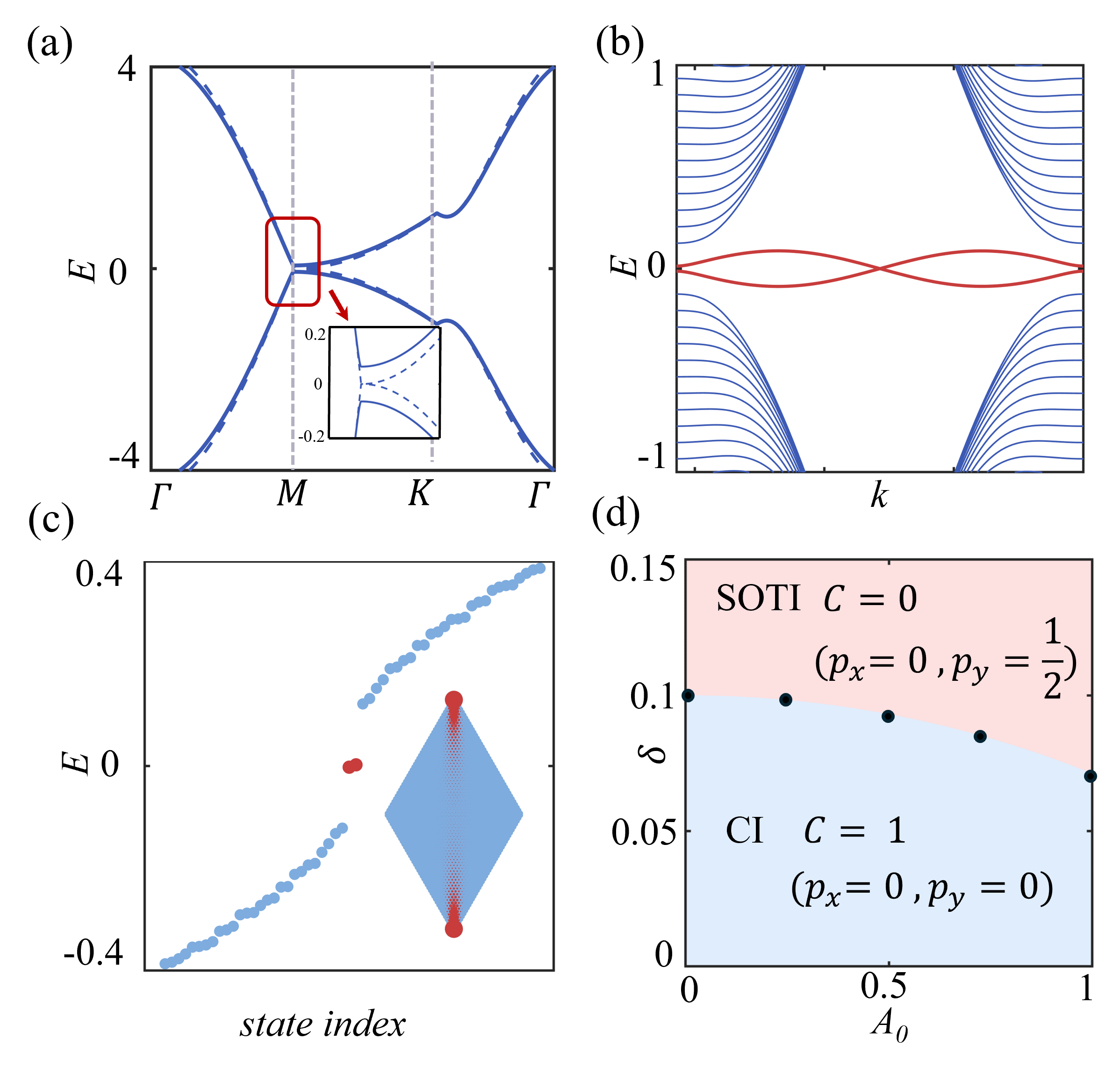}
    \caption{Semi-Dirac regime and Floquet second-order topology. (a) Bulk dispersion near the Dirac-merging point: two Dirac cones merge at $M$
to form a semi-Dirac spectrum (linear in one direction and quadratic in the
other). The driven system opens an anisotropic gap (inset). Parameters: $t_0=1,\omega=3,A_0=0.5,\phi=\frac{\pi}{3},\theta =\frac{\pi}{2},\delta=0.1$, and $\beta=7.8$. (b) Ribbon quasienergy spectrum under oblique incidence: edge states become gapped and form isolated edge-localized bands separated from the bulk continuum. (c) Rhombic finite geometry showing in-gap corner modes (two states) localized at the two acute corners, consistent with a Floquet SOTI with gapped edges. (d) Phase diagram versus strain and drive strength at fixed oblique incidence (e.g., $\phi=\pi/3$), labeled by $(C,p_y)$: Chern insulator (CI) with $(C,p_y)=(1,0)$ and Floquet SOTI with $(C,p_y)=(0,1/2)$.}
    \label{fig3}
\end{figure}

\emph{Floquet second-order topological insulator}---We now focus on the strain-tuned Dirac-merging regime, where uniaxial strain not only shifts the Dirac points but also strongly renormalizes the band velocities and drives the spectrum toward a semi-Dirac form with mixed linear--quadratic dispersion [Fig.~\ref{fig3}(a)]. At the critical strain, two Dirac points with opposite winding merge at the $M$ point. In this regime, the net Berry curvature and thus the Chern response are not the most natural diagnostics of the driven phase. Instead, the key ingredient is that oblique-incidence (elliptically projected) driving can open a bulk gap while simultaneously generating orientation-dependent boundary masses. In a ribbon geometry, this manifests as gapped boundary bands that detach from the bulk manifolds, forming energetically isolated edge-localized bands [Fig.~\ref{fig3}(b)]. In a finite sample, sign changes of the edge mass on adjacent edges produce domain walls at corners, yielding in-gap corner modes [Fig.~\ref{fig3}(c)].

The stability and diagnosis of this Floquet SOTI rely on an antiunitary crystalline symmetry that survives the combined action of strain and oblique driving. While $\mathcal{T}$ and ${C}_{2x}$ are individually broken, their product remains a symmetry of ${\cal H}_{\rm eff}$,
${\mathcal{M}}={{C}}_{2x}{\mathcal{T}}$. For spinless fermions ${\mathcal{T}}=\mathcal{K}$ (complex conjugation), and ${C}_{2x}$ exchanges the two sublattices, represented by ${{C}}_{2x}=\sigma_x$ in the $(A,B)$ basis. 
Accordingly, we characterize the second-order topology by the polarization indices along the $x_i$ axis~\cite{Ezawa2018PRL},
\begin{equation}
p_i=\frac{1}{S}\int_{\rm BZ} d^2k\,\mathcal{A}_i(\bm k)\quad{\rm mod}\;1,
\end{equation}
where $\mathcal{A}_i(\bm k)=-i\langle u(\bm k)|\partial_{k_i}|u(\bm k)\rangle$ with $|u(\bm k)\rangle$ the occupied Floquet eigenstate, and $S$ is the area of BZ. The symmetry ${\mathcal{M}}$ quantizes the transverse polarization $p_y$ to $0$ or $1/2$ (mod 1). Physically, $p_i=0$ corresponds to Wannier charge centers located at the unit-cell center along direction $i$, whereas $p_i=1/2$ indicates a shift to the unit-cell boundary midpoint. In the Floquet SOTI phase, the drive-induced anisotropic mass pins the Wannier centers such that $(p_x,p_y)=(0,\tfrac{1}{2})$, signaling an obstructed atomic limit distinct from a trivial atomic insulator and guaranteeing corner-localized in-gap modes in a finite geometry.

Finally, by evaluating both $\mathcal{C}$ and $(p_x,p_y)$ we obtain the phase diagram versus strain magnitude $\delta$ and drive strength $A_0$ at fixed oblique incidence $\phi=\pi/3$ [Fig.~\ref{fig3}(d)]. For moderate anisotropy ($t_1<2t_2$), the driven system is a first-order Floquet topological insulator characterized by a nonzero Chern number and chiral edge modes. Beyond the Dirac-merging threshold ($t_1>2t_2$), the Chern number becomes trivial while the symmetry-quantized polarization locks to $p_y=1/2$, identifying a Floquet second-order topological insulator with gapped edges and robust corner states.

\begin{figure}[t!]
    \centering
    \includegraphics[width=1\linewidth]{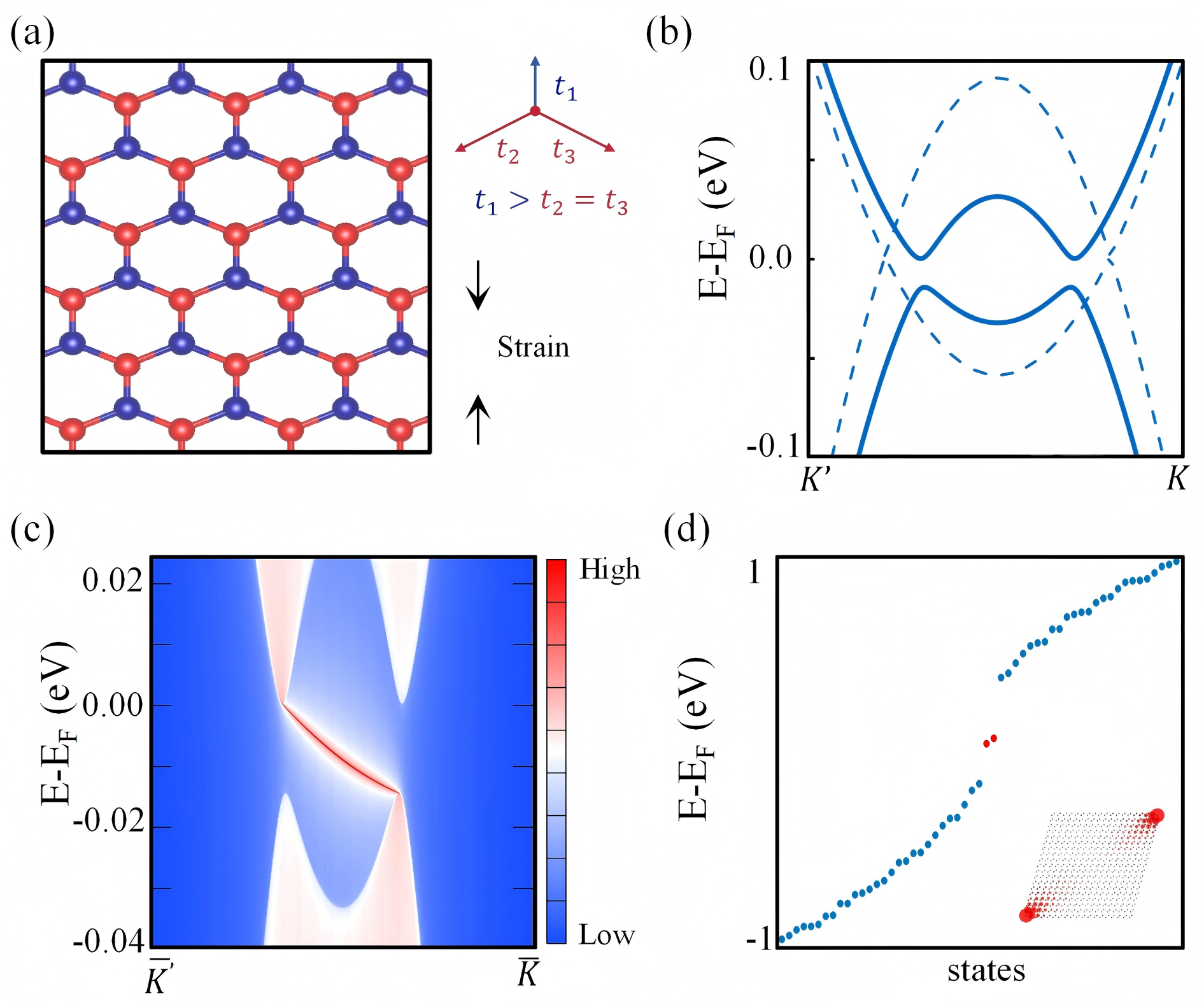}
    \caption{First-principles-informed realization in strained graphene.
(a) Distorted graphene lattice used in DFT/Wannier modeling, yielding an
effective hopping hierarchy $t_1>t_2=t_3$. (b) Wannier tight-binding bands without driving (gray dashed) and with driving (red) from $H_{\rm eff}$ at the indicated $\hbar\omega$ and $\mathcal{A}_0$. (c) Ribbon spectrum in the intermediate-intensity window showing a chiral edge mode consistent with $C=1$.
(d) Nanodisk spectrum at larger intensity displaying two in-gap corner modes;
their splitting decreases with increasing disk size, consistent with
exponentially localized corner states.}
    \label{fig4}
\end{figure}

\emph{First-principles-informed material realization.}---To substantiate the model analysis with a realistic material description, we consider monolayer graphene and perform first-principles calculations followed by a Wannier-based TB construction, which enables an efficient implementation of the Floquet drive in real space. Our goal is to verify, within a material-derived Hamiltonian, the same sequence of photoinduced phases predicted by the strained honeycomb model: a gapped Floquet Chern phase at intermediate intensity and a higher-order phase with corner-localized modes at stronger driving.

Density functional theory (DFT) calculations~\cite{Hohenberg1964,Kohn1965} are carried out using the Vienna \textit{ab initio} simulation package (VASP)~\cite{Kresse1996} with the projector augmented-wave (PAW) method and the Perdew--Burke--Ernzerhof (PBE) exchange-correlation functional~\cite{Perdew1996}. The plane-wave cutoff is set to 520~eV and the BZ is sampled on a $15\times15\times1$ Monkhorst--Pack grid. To emulate the hopping anisotropy required to access the Dirac-merging regime in our model, we apply a uniaxial lattice distortion such that the angle between the primitive vectors is tuned from $60^\circ$ to $74^\circ$ [Fig.~\ref{fig4}(a)], corresponding to an effective strain of $\epsilon\approx 8.09\%$, and producing an effective hierarchy $t_1>t_2=t_3$ at the TB level.

From the DFT band structure we construct a real-space TB Hamiltonian using maximally localized Wannier functions (MLWFs)~\cite{WU2018405}. We verify that the MLWF TB bands faithfully reproduce the DFT dispersion in the energy window of interest, ensuring that the subsequent Floquet analysis is performed on a quantitatively reliable low-energy model. The optical drive is then introduced via the Peierls substitution on the Wannier TB hoppings, and we extract the corresponding Floquet effective bands in the off-resonant regime.

As the light intensity increases, the Dirac points shift and gap out due to the drive-induced time-reversal breaking [Fig.~\ref{fig4}(b)]. In an intermediate-intensity window, the driven band structure exhibits a band inversion and the system enters a Floquet Chern insulating phase with Chern number $\mathcal{C}=1$, accompanied by a chiral edge mode in a ribbon geometry [Fig.~\ref{fig4}(c)]. This behavior is fully consistent with the effective Floquet Hamiltonian derived in the model analysis. Upon further increasing the intensity, the inverted gap is removed and the bands become fully separated; correspondingly, the first-order topology becomes trivial and the Chern number returns to zero. Importantly, although $\mathcal{C}=0$ in this regime, our model predicts that a higher-order phase can still be realized due to the emergence of orientation-dependent boundary masses. To test this material-realistic scenario, we compute the spectrum of a finite graphene nanodisk [Fig.~\ref{fig4}(d)] and observe two robust in-gap modes localized at the two acute corners. Their finite-size splitting decreases as the disk size increases, consistent with exponentially localized corner states in the thermodynamic limit. Together, these results establish a complete topological evolution in driven strained graphene: Dirac semimetal $\rightarrow$ Floquet Chern insulator $\rightarrow$ Floquet higher-order topological insulator.

\emph{Conclusion.}---We have proposed a practical route to Floquet higher-order topology in graphene by combining uniaxial strain with off-resonant CPL at a tunable incidence angle. We emphasize that this work is a theoretical proposal based on the simplified high-frequency approximation, and recent experiments demonstrate the growing feasibility of Floquet-driven phases in graphene and related materials~\cite{Choi2025,Merboldt2025,Wang2026}. Strain tunes graphene toward the Dirac-merging (semi-Dirac) regime, where the light-induced mass becomes strongly anisotropic. Under oblique incidence, the in-plane projection of CPL is effectively elliptically polarized, and in conjunction with strain it generates orientation-dependent boundary masses that gap the edges while enforcing mass sign changes between adjacent edges. Consequently, the driven system can be switched from a first-order Floquet Chern insulator with chiral edge modes to a Floquet SOTI with gapped edges and robust in-gap corner states. The phase boundaries are determined by the Chern number together with a symmetry-quantized polarization/Wannier-center invariant protected by an antiunitary crystalline symmetry. First-principles-informed Wannier-Floquet calculations corroborate the predicted phase sequence in strained graphene nanostructures, demonstrating that driven strained graphene provides a realistic and tunable platform for realizing and diagnosing Floquet higher-order topological phases.
As in other periodically driven electronic systems, heating ultimately limits the lifetime of the Floquet-engineered phase; accordingly, our results should be understood as describing a prethermal off-resonant regime, in which resonant absorption is suppressed and the effective Hamiltonian captures the leading-order topology over experimentally relevant timescales.

\emph{Acknowledgments.}---This work was supported by National Key Research and Development Program of the Ministry of Science and Technology of China (Grant No. 2025YFA1411303), the National Natural Science Foundation of China (NSFC, Grants No. 92565103, No. 12474151, No. 12222402, and No. 12547101), the Natural Science Foundation of Chongqing (Grant No. CSTB2025NSCQ-LZX0010), Beijing National Laboratory for Condensed Matter Physics (No. 2024BNLCMPKF025), and the Fundamental Research Funds for the Central Universities (Grant No. 2025CDJIAISYB-032).

\appendix
\section{Appendix: Low-energy Dirac-merging form}
In unstrained graphene, expanding the Bloch Hamiltonian around the $K$ point yields the standard linear Dirac form
\begin{equation}
{\cal H}_{\rm low}(\bm k)\simeq -\frac{3}{2}t\,k_x\sigma_x+\frac{3}{2}t\,k_y\sigma_y+{\cal O}(\bm k^2),
\end{equation}
in the present lattice units (bond length set to unity). Under uniaxial strain with $t_2=t_3\equiv t$ and $t_1>t$, the Dirac points move toward the $M$ point and the dispersion becomes strongly anisotropic. At the Dirac-merging threshold $t_1=2t$, the low-energy expansion about $M$ takes the semi-Dirac form
\begin{equation}
{\cal H}_{\rm SD}(\bm k)\simeq \frac{k_x^2}{2m}\,\sigma_x+v\,k_y\,\sigma_y,
\qquad
m=\frac{2}{3t},\quad v=\frac{3t}{2},
\end{equation}
which is quadratic along $k_x$ and linear along $k_y$. This mixed dispersion enhances the directional sensitivity of the Floquet mass near the critical regime and provides the microscopic basis for the edge-dependent boundary gaps and corner-domain-wall modes discussed in the main text.

\bibliography{bibfile2.bib}

\begin{thebibliography}{99}%
\makeatletter
\providecommand \@ifxundefined [1]{%
 \@ifx{#1\undefined}
}%
\providecommand \@ifnum [1]{%
 \ifnum #1\expandafter \@firstoftwo
 \else \expandafter \@secondoftwo
 \fi
}%
\providecommand \@ifx [1]{%
 \ifx #1\expandafter \@firstoftwo
 \else \expandafter \@secondoftwo
 \fi
}%
\providecommand \natexlab [1]{#1}%
\providecommand \enquote  [1]{``#1''}%
\providecommand \bibnamefont  [1]{#1}%
\providecommand \bibfnamefont [1]{#1}%
\providecommand \citenamefont [1]{#1}%
\providecommand \href@noop [0]{\@secondoftwo}%
\providecommand \href [0]{\begingroup \@sanitize@url \@href}%
\providecommand \@href[1]{\@@startlink{#1}\@@href}%
\providecommand \@@href[1]{\endgroup#1\@@endlink}%
\providecommand \@sanitize@url [0]{\catcode `\\12\catcode `\$12\catcode
  `\&12\catcode `\#12\catcode `\^12\catcode `\_12\catcode `\%12\relax}%
\providecommand \@@startlink[1]{}%
\providecommand \@@endlink[0]{}%
\providecommand \url  [0]{\begingroup\@sanitize@url \@url }%
\providecommand \@url [1]{\endgroup\@href {#1}{\urlprefix }}%
\providecommand \urlprefix  [0]{URL }%
\providecommand \Eprint [0]{\href }%
\providecommand \doibase [0]{https://doi.org/}%
\providecommand \selectlanguage [0]{\@gobble}%
\providecommand \bibinfo  [0]{\@secondoftwo}%
\providecommand \bibfield  [0]{\@secondoftwo}%
\providecommand \translation [1]{[#1]}%
\providecommand \BibitemOpen [0]{}%
\providecommand \bibitemStop [0]{}%
\providecommand \bibitemNoStop [0]{.\EOS\space}%
\providecommand \EOS [0]{\spacefactor3000\relax}%
\providecommand \BibitemShut  [1]{\csname bibitem#1\endcsname}%
\let\auto@bib@innerbib\@empty
\bibitem [{\citenamefont {Vafek}\ and\ \citenamefont
  {Vishwanath}(2014)}]{Dirac-review}%
  \BibitemOpen
  \bibfield  {author} {\bibinfo {author} {\bibfnamefont {O.}~\bibnamefont
  {Vafek}}\ and\ \bibinfo {author} {\bibfnamefont {A.}~\bibnamefont
  {Vishwanath}},\ }\bibfield  {title} {\bibinfo {title} {Dirac fermions in
  solids: From high-{T}$_c$ cuprates and graphene to topological insulators and
  {W}eyl semimetals},\ }\href
  {https://doi.org/10.1146/annurev-conmatphys-031113-133841} {\bibfield
  {journal} {\bibinfo  {journal} {Annu. Rev. Condens. Matter Phys.}\ }\textbf
  {\bibinfo {volume} {5}},\ \bibinfo {pages} {83} (\bibinfo {year}
  {2014})}\BibitemShut {NoStop}%
\bibitem [{\citenamefont {Castro~Neto}\ \emph {et~al.}(2009)\citenamefont
  {Castro~Neto}, \citenamefont {Guinea}, \citenamefont {Peres}, \citenamefont
  {Novoselov},\ and\ \citenamefont {Geim}}]{Graphene-Review}%
  \BibitemOpen
  \bibfield  {author} {\bibinfo {author} {\bibfnamefont {A.~H.}\ \bibnamefont
  {Castro~Neto}}, \bibinfo {author} {\bibfnamefont {F.}~\bibnamefont {Guinea}},
  \bibinfo {author} {\bibfnamefont {N.~M.~R.}\ \bibnamefont {Peres}}, \bibinfo
  {author} {\bibfnamefont {K.~S.}\ \bibnamefont {Novoselov}},\ and\ \bibinfo
  {author} {\bibfnamefont {A.~K.}\ \bibnamefont {Geim}},\ }\bibfield  {title}
  {\bibinfo {title} {The electronic properties of graphene},\ }\href
  {https://doi.org/10.1103/RevModPhys.81.109} {\bibfield  {journal} {\bibinfo
  {journal} {Rev. Mod. Phys.}\ }\textbf {\bibinfo {volume} {81}},\ \bibinfo
  {pages} {109} (\bibinfo {year} {2009})}\BibitemShut {NoStop}%
\bibitem [{\citenamefont {Haldane}(1988)}]{Haldane1988}%
  \BibitemOpen
  \bibfield  {author} {\bibinfo {author} {\bibfnamefont {F.~D.~M.}\
  \bibnamefont {Haldane}},\ }\bibfield  {title} {\bibinfo {title} {Model for a
  quantum {H}all effect without {L}andau levels: Condensed-matter realization
  of the ``parity anomaly"},\ }\href
  {https://doi.org/10.1103/PhysRevLett.61.2015} {\bibfield  {journal} {\bibinfo
   {journal} {Phys. Rev. Lett.}\ }\textbf {\bibinfo {volume} {61}},\ \bibinfo
  {pages} {2015} (\bibinfo {year} {1988})}\BibitemShut {NoStop}%
\bibitem [{\citenamefont {Kane}\ and\ \citenamefont {Mele}(2005)}]{KMmodel}%
  \BibitemOpen
  \bibfield  {author} {\bibinfo {author} {\bibfnamefont {C.~L.}\ \bibnamefont
  {Kane}}\ and\ \bibinfo {author} {\bibfnamefont {E.~J.}\ \bibnamefont
  {Mele}},\ }\bibfield  {title} {\bibinfo {title} {Quantum spin hall effect in
  graphene},\ }\href {https://doi.org/10.1103/PhysRevLett.95.226801} {\bibfield
   {journal} {\bibinfo  {journal} {Phys. Rev. Lett.}\ }\textbf {\bibinfo
  {volume} {95}},\ \bibinfo {pages} {226801} (\bibinfo {year}
  {2005})}\BibitemShut {NoStop}%
\bibitem [{\citenamefont {Cayssol}\ \emph {et~al.}(2013)\citenamefont
  {Cayssol}, \citenamefont {D{\'o}ra}, \citenamefont {Simon},\ and\
  \citenamefont {Moessner}}]{pssr.201206451}%
  \BibitemOpen
  \bibfield  {author} {\bibinfo {author} {\bibfnamefont {J.}~\bibnamefont
  {Cayssol}}, \bibinfo {author} {\bibfnamefont {B.}~\bibnamefont {D{\'o}ra}},
  \bibinfo {author} {\bibfnamefont {F.}~\bibnamefont {Simon}},\ and\ \bibinfo
  {author} {\bibfnamefont {R.}~\bibnamefont {Moessner}},\ }\bibfield  {title}
  {\bibinfo {title} {Floquet topological insulators},\ }\href
  {https://doi.org/10.1002/pssr.201206451} {\bibfield  {journal} {\bibinfo
  {journal} {Phys. Status Solidi RRL}\ }\textbf {\bibinfo {volume} {7}},\
  \bibinfo {pages} {101} (\bibinfo {year} {2013})}\BibitemShut {NoStop}%
\bibitem [{\citenamefont {Oka}\ and\ \citenamefont
  {Kitamura}(2019)}]{annurev-conmatphys-031218-013423}%
  \BibitemOpen
  \bibfield  {author} {\bibinfo {author} {\bibfnamefont {T.}~\bibnamefont
  {Oka}}\ and\ \bibinfo {author} {\bibfnamefont {S.}~\bibnamefont {Kitamura}},\
  }\bibfield  {title} {\bibinfo {title} {Floquet engineering of quantum
  materials},\ }\href
  {https://doi.org/10.1146/annurev-conmatphys-031218-013423} {\bibfield
  {journal} {\bibinfo  {journal} {Annu. Rev. Condens. Matter Phys.}\ }\textbf
  {\bibinfo {volume} {10}},\ \bibinfo {pages} {387} (\bibinfo {year}
  {2019})}\BibitemShut {NoStop}%
\bibitem [{\citenamefont {Harper}\ \emph {et~al.}(2020)\citenamefont {Harper},
  \citenamefont {Roy}, \citenamefont {Rudner},\ and\ \citenamefont
  {Sondhi}}]{harper-annurev-conmatphys}%
  \BibitemOpen
  \bibfield  {author} {\bibinfo {author} {\bibfnamefont {F.}~\bibnamefont
  {Harper}}, \bibinfo {author} {\bibfnamefont {R.}~\bibnamefont {Roy}},
  \bibinfo {author} {\bibfnamefont {M.~S.}\ \bibnamefont {Rudner}},\ and\
  \bibinfo {author} {\bibfnamefont {S.}~\bibnamefont {Sondhi}},\ }\bibfield
  {title} {\bibinfo {title} {Topology and broken symmetry in {F}loquet
  systems},\ }\href {https://doi.org/10.1146/annurev-conmatphys-031218-013721}
  {\bibfield  {journal} {\bibinfo  {journal} {Annu. Rev. Condens. Matter
  Phys.}\ }\textbf {\bibinfo {volume} {11}},\ \bibinfo {pages} {345} (\bibinfo
  {year} {2020})}\BibitemShut {NoStop}%
\bibitem [{\citenamefont {Rudner}\ and\ \citenamefont
  {Lindner}(2020)}]{rudner2020band}%
  \BibitemOpen
  \bibfield  {author} {\bibinfo {author} {\bibfnamefont {M.~S.}\ \bibnamefont
  {Rudner}}\ and\ \bibinfo {author} {\bibfnamefont {N.~H.}\ \bibnamefont
  {Lindner}},\ }\bibfield  {title} {\bibinfo {title} {Band structure
  engineering and non-equilibrium dynamics in {F}loquet topological
  insulators},\ }\href {https://doi.org/10.1038/s42254-020-0170-z} {\bibfield
  {journal} {\bibinfo  {journal} {Nat. Rev. Phys.}\ }\textbf {\bibinfo {volume}
  {2}},\ \bibinfo {pages} {229} (\bibinfo {year} {2020})}\BibitemShut {NoStop}%
\bibitem [{\citenamefont {Bao}\ \emph {et~al.}(2022)\citenamefont {Bao},
  \citenamefont {Tang}, \citenamefont {Sun},\ and\ \citenamefont
  {Zhou}}]{bao2022light}%
  \BibitemOpen
  \bibfield  {author} {\bibinfo {author} {\bibfnamefont {C.}~\bibnamefont
  {Bao}}, \bibinfo {author} {\bibfnamefont {P.}~\bibnamefont {Tang}}, \bibinfo
  {author} {\bibfnamefont {D.}~\bibnamefont {Sun}},\ and\ \bibinfo {author}
  {\bibfnamefont {S.}~\bibnamefont {Zhou}},\ }\bibfield  {title} {\bibinfo
  {title} {Light-induced emergent phenomena in 2{D} materials and topological
  materials},\ }\href {https://doi.org/10.1038/s42254-021-00388-1} {\bibfield
  {journal} {\bibinfo  {journal} {Nat. Rev. Phys.}\ }\textbf {\bibinfo {volume}
  {4}},\ \bibinfo {pages} {33} (\bibinfo {year} {2022})}\BibitemShut {NoStop}%
\bibitem [{\citenamefont {Oka}\ and\ \citenamefont
  {Aoki}(2009)}]{PhysRevB.79.081406}%
  \BibitemOpen
  \bibfield  {author} {\bibinfo {author} {\bibfnamefont {T.}~\bibnamefont
  {Oka}}\ and\ \bibinfo {author} {\bibfnamefont {H.}~\bibnamefont {Aoki}},\
  }\bibfield  {title} {\bibinfo {title} {Photovoltaic {H}all effect in
  graphene},\ }\href {https://doi.org/10.1103/PhysRevB.79.081406} {\bibfield
  {journal} {\bibinfo  {journal} {Phys. Rev. B}\ }\textbf {\bibinfo {volume}
  {79}},\ \bibinfo {pages} {081406} (\bibinfo {year} {2009})}\BibitemShut
  {NoStop}%
\bibitem [{\citenamefont {Inoue}\ and\ \citenamefont
  {Tanaka}(2010)}]{PhysRevLett.105.017401}%
  \BibitemOpen
  \bibfield  {author} {\bibinfo {author} {\bibfnamefont {J.-i.}\ \bibnamefont
  {Inoue}}\ and\ \bibinfo {author} {\bibfnamefont {A.}~\bibnamefont {Tanaka}},\
  }\bibfield  {title} {\bibinfo {title} {Photoinduced transition between
  conventional and topological insulators in two-dimensional electronic
  systems},\ }\href {https://doi.org/10.1103/PhysRevLett.105.017401} {\bibfield
   {journal} {\bibinfo  {journal} {Phys. Rev. Lett.}\ }\textbf {\bibinfo
  {volume} {105}},\ \bibinfo {pages} {017401} (\bibinfo {year}
  {2010})}\BibitemShut {NoStop}%
\bibitem [{\citenamefont {Kitagawa}\ \emph
  {et~al.}(2010{\natexlab{a}})\citenamefont {Kitagawa}, \citenamefont {Berg},
  \citenamefont {Rudner},\ and\ \citenamefont {Demler}}]{PhysRevB.82.235114}%
  \BibitemOpen
  \bibfield  {author} {\bibinfo {author} {\bibfnamefont {T.}~\bibnamefont
  {Kitagawa}}, \bibinfo {author} {\bibfnamefont {E.}~\bibnamefont {Berg}},
  \bibinfo {author} {\bibfnamefont {M.}~\bibnamefont {Rudner}},\ and\ \bibinfo
  {author} {\bibfnamefont {E.}~\bibnamefont {Demler}},\ }\bibfield  {title}
  {\bibinfo {title} {Topological characterization of periodically driven
  quantum systems},\ }\href {https://doi.org/10.1103/PhysRevB.82.235114}
  {\bibfield  {journal} {\bibinfo  {journal} {Phys. Rev. B}\ }\textbf {\bibinfo
  {volume} {82}},\ \bibinfo {pages} {235114} (\bibinfo {year}
  {2010}{\natexlab{a}})}\BibitemShut {NoStop}%
\bibitem [{\citenamefont {Gu}\ \emph {et~al.}(2011)\citenamefont {Gu},
  \citenamefont {Fertig}, \citenamefont {Arovas},\ and\ \citenamefont
  {Auerbach}}]{PhysRevLett.107.216601}%
  \BibitemOpen
  \bibfield  {author} {\bibinfo {author} {\bibfnamefont {Z.}~\bibnamefont
  {Gu}}, \bibinfo {author} {\bibfnamefont {H.~A.}\ \bibnamefont {Fertig}},
  \bibinfo {author} {\bibfnamefont {D.~P.}\ \bibnamefont {Arovas}},\ and\
  \bibinfo {author} {\bibfnamefont {A.}~\bibnamefont {Auerbach}},\ }\bibfield
  {title} {\bibinfo {title} {Floquet spectrum and transport through an
  irradiated graphene ribbon},\ }\href
  {https://doi.org/10.1103/PhysRevLett.107.216601} {\bibfield  {journal}
  {\bibinfo  {journal} {Phys. Rev. Lett.}\ }\textbf {\bibinfo {volume} {107}},\
  \bibinfo {pages} {216601} (\bibinfo {year} {2011})}\BibitemShut {NoStop}%
\bibitem [{\citenamefont {Kitagawa}\ \emph {et~al.}(2011)\citenamefont
  {Kitagawa}, \citenamefont {Oka}, \citenamefont {Brataas}, \citenamefont
  {Fu},\ and\ \citenamefont {Demler}}]{PhysRevB.84.235108}%
  \BibitemOpen
  \bibfield  {author} {\bibinfo {author} {\bibfnamefont {T.}~\bibnamefont
  {Kitagawa}}, \bibinfo {author} {\bibfnamefont {T.}~\bibnamefont {Oka}},
  \bibinfo {author} {\bibfnamefont {A.}~\bibnamefont {Brataas}}, \bibinfo
  {author} {\bibfnamefont {L.}~\bibnamefont {Fu}},\ and\ \bibinfo {author}
  {\bibfnamefont {E.}~\bibnamefont {Demler}},\ }\bibfield  {title} {\bibinfo
  {title} {Transport properties of nonequilibrium systems under the application
  of light: Photoinduced quantum {H}all insulators without {L}andau levels},\
  }\href {https://doi.org/10.1103/PhysRevB.84.235108} {\bibfield  {journal}
  {\bibinfo  {journal} {Phys. Rev. B}\ }\textbf {\bibinfo {volume} {84}},\
  \bibinfo {pages} {235108} (\bibinfo {year} {2011})}\BibitemShut {NoStop}%
\bibitem [{\citenamefont {Delplace}\ \emph {et~al.}(2013)\citenamefont
  {Delplace}, \citenamefont {G\'omez-Le\'on},\ and\ \citenamefont
  {Platero}}]{PhysRevB.88.245422}%
  \BibitemOpen
  \bibfield  {author} {\bibinfo {author} {\bibfnamefont {P.}~\bibnamefont
  {Delplace}}, \bibinfo {author} {\bibfnamefont {A.}~\bibnamefont
  {G\'omez-Le\'on}},\ and\ \bibinfo {author} {\bibfnamefont {G.}~\bibnamefont
  {Platero}},\ }\bibfield  {title} {\bibinfo {title} {Merging of {D}irac points
  and {F}loquet topological transitions in ac-driven graphene},\ }\href
  {https://doi.org/10.1103/PhysRevB.88.245422} {\bibfield  {journal} {\bibinfo
  {journal} {Phys. Rev. B}\ }\textbf {\bibinfo {volume} {88}},\ \bibinfo
  {pages} {245422} (\bibinfo {year} {2013})}\BibitemShut {NoStop}%
\bibitem [{\citenamefont {Foa~Torres}\ \emph {et~al.}(2014)\citenamefont
  {Foa~Torres}, \citenamefont {Perez-Piskunow}, \citenamefont {Balseiro},\ and\
  \citenamefont {Usaj}}]{PhysRevLett.113.266801}%
  \BibitemOpen
  \bibfield  {author} {\bibinfo {author} {\bibfnamefont {L.~E.~F.}\
  \bibnamefont {Foa~Torres}}, \bibinfo {author} {\bibfnamefont {P.~M.}\
  \bibnamefont {Perez-Piskunow}}, \bibinfo {author} {\bibfnamefont {C.~A.}\
  \bibnamefont {Balseiro}},\ and\ \bibinfo {author} {\bibfnamefont
  {G.}~\bibnamefont {Usaj}},\ }\bibfield  {title} {\bibinfo {title}
  {Multiterminal conductance of a {F}loquet topological insulator},\ }\href
  {https://doi.org/10.1103/PhysRevLett.113.266801} {\bibfield  {journal}
  {\bibinfo  {journal} {Phys. Rev. Lett.}\ }\textbf {\bibinfo {volume} {113}},\
  \bibinfo {pages} {266801} (\bibinfo {year} {2014})}\BibitemShut {NoStop}%
\bibitem [{\citenamefont {Usaj}\ \emph {et~al.}(2014)\citenamefont {Usaj},
  \citenamefont {Perez-Piskunow}, \citenamefont {Foa~Torres},\ and\
  \citenamefont {Balseiro}}]{PhysRevB.90.115423}%
  \BibitemOpen
  \bibfield  {author} {\bibinfo {author} {\bibfnamefont {G.}~\bibnamefont
  {Usaj}}, \bibinfo {author} {\bibfnamefont {P.~M.}\ \bibnamefont
  {Perez-Piskunow}}, \bibinfo {author} {\bibfnamefont {L.~E.~F.}\ \bibnamefont
  {Foa~Torres}},\ and\ \bibinfo {author} {\bibfnamefont {C.~A.}\ \bibnamefont
  {Balseiro}},\ }\bibfield  {title} {\bibinfo {title} {Irradiated graphene as a
  tunable {F}loquet topological insulator},\ }\href
  {https://doi.org/10.1103/PhysRevB.90.115423} {\bibfield  {journal} {\bibinfo
  {journal} {Phys. Rev. B}\ }\textbf {\bibinfo {volume} {90}},\ \bibinfo
  {pages} {115423} (\bibinfo {year} {2014})}\BibitemShut {NoStop}%
\bibitem [{\citenamefont {Lindner}\ \emph {et~al.}(2011)\citenamefont
  {Lindner}, \citenamefont {Refael},\ and\ \citenamefont
  {Galitski}}]{lindner2011floquet}%
  \BibitemOpen
  \bibfield  {author} {\bibinfo {author} {\bibfnamefont {N.~H.}\ \bibnamefont
  {Lindner}}, \bibinfo {author} {\bibfnamefont {G.}~\bibnamefont {Refael}},\
  and\ \bibinfo {author} {\bibfnamefont {V.}~\bibnamefont {Galitski}},\
  }\bibfield  {title} {\bibinfo {title} {Floquet topological insulator in
  semiconductor quantum wells},\ }\href {https://doi.org/10.1038/nphys1926}
  {\bibfield  {journal} {\bibinfo  {journal} {Nat. Phys.}\ }\textbf {\bibinfo
  {volume} {7}},\ \bibinfo {pages} {490} (\bibinfo {year} {2011})}\BibitemShut
  {NoStop}%
\bibitem [{\citenamefont {Kitagawa}\ \emph
  {et~al.}(2010{\natexlab{b}})\citenamefont {Kitagawa}, \citenamefont {Rudner},
  \citenamefont {Berg},\ and\ \citenamefont {Demler}}]{PhysRevA.82.033429}%
  \BibitemOpen
  \bibfield  {author} {\bibinfo {author} {\bibfnamefont {T.}~\bibnamefont
  {Kitagawa}}, \bibinfo {author} {\bibfnamefont {M.~S.}\ \bibnamefont
  {Rudner}}, \bibinfo {author} {\bibfnamefont {E.}~\bibnamefont {Berg}},\ and\
  \bibinfo {author} {\bibfnamefont {E.}~\bibnamefont {Demler}},\ }\bibfield
  {title} {\bibinfo {title} {Exploring topological phases with quantum walks},\
  }\href {https://doi.org/10.1103/PhysRevA.82.033429} {\bibfield  {journal}
  {\bibinfo  {journal} {Phys. Rev. A}\ }\textbf {\bibinfo {volume} {82}},\
  \bibinfo {pages} {033429} (\bibinfo {year} {2010}{\natexlab{b}})}\BibitemShut
  {NoStop}%
\bibitem [{\citenamefont {Lindner}\ \emph {et~al.}(2013)\citenamefont
  {Lindner}, \citenamefont {Bergman}, \citenamefont {Refael},\ and\
  \citenamefont {Galitski}}]{PhysRevB.87.235131}%
  \BibitemOpen
  \bibfield  {author} {\bibinfo {author} {\bibfnamefont {N.~H.}\ \bibnamefont
  {Lindner}}, \bibinfo {author} {\bibfnamefont {D.~L.}\ \bibnamefont
  {Bergman}}, \bibinfo {author} {\bibfnamefont {G.}~\bibnamefont {Refael}},\
  and\ \bibinfo {author} {\bibfnamefont {V.}~\bibnamefont {Galitski}},\
  }\bibfield  {title} {\bibinfo {title} {Topological {F}loquet spectrum in
  three dimensions via a two-photon resonance},\ }\href
  {https://doi.org/10.1103/PhysRevB.87.235131} {\bibfield  {journal} {\bibinfo
  {journal} {Phys. Rev. B}\ }\textbf {\bibinfo {volume} {87}},\ \bibinfo
  {pages} {235131} (\bibinfo {year} {2013})}\BibitemShut {NoStop}%
\bibitem [{\citenamefont {Jiang}\ \emph {et~al.}(2011)\citenamefont {Jiang},
  \citenamefont {Kitagawa}, \citenamefont {Alicea}, \citenamefont {Akhmerov},
  \citenamefont {Pekker}, \citenamefont {Refael}, \citenamefont {Cirac},
  \citenamefont {Demler}, \citenamefont {Lukin},\ and\ \citenamefont
  {Zoller}}]{PhysRevLett.106.220402}%
  \BibitemOpen
  \bibfield  {author} {\bibinfo {author} {\bibfnamefont {L.}~\bibnamefont
  {Jiang}}, \bibinfo {author} {\bibfnamefont {T.}~\bibnamefont {Kitagawa}},
  \bibinfo {author} {\bibfnamefont {J.}~\bibnamefont {Alicea}}, \bibinfo
  {author} {\bibfnamefont {A.~R.}\ \bibnamefont {Akhmerov}}, \bibinfo {author}
  {\bibfnamefont {D.}~\bibnamefont {Pekker}}, \bibinfo {author} {\bibfnamefont
  {G.}~\bibnamefont {Refael}}, \bibinfo {author} {\bibfnamefont {J.~I.}\
  \bibnamefont {Cirac}}, \bibinfo {author} {\bibfnamefont {E.}~\bibnamefont
  {Demler}}, \bibinfo {author} {\bibfnamefont {M.~D.}\ \bibnamefont {Lukin}},\
  and\ \bibinfo {author} {\bibfnamefont {P.}~\bibnamefont {Zoller}},\
  }\bibfield  {title} {\bibinfo {title} {Majorana fermions in equilibrium and
  in driven cold-atom quantum wires},\ }\href
  {https://doi.org/10.1103/PhysRevLett.106.220402} {\bibfield  {journal}
  {\bibinfo  {journal} {Phys. Rev. Lett.}\ }\textbf {\bibinfo {volume} {106}},\
  \bibinfo {pages} {220402} (\bibinfo {year} {2011})}\BibitemShut {NoStop}%
\bibitem [{\citenamefont {Rudner}\ \emph {et~al.}(2013)\citenamefont {Rudner},
  \citenamefont {Lindner}, \citenamefont {Berg},\ and\ \citenamefont
  {Levin}}]{PhysRevX.3.031005}%
  \BibitemOpen
  \bibfield  {author} {\bibinfo {author} {\bibfnamefont {M.~S.}\ \bibnamefont
  {Rudner}}, \bibinfo {author} {\bibfnamefont {N.~H.}\ \bibnamefont {Lindner}},
  \bibinfo {author} {\bibfnamefont {E.}~\bibnamefont {Berg}},\ and\ \bibinfo
  {author} {\bibfnamefont {M.}~\bibnamefont {Levin}},\ }\bibfield  {title}
  {\bibinfo {title} {Anomalous edge states and the bulk-edge correspondence for
  periodically driven two-dimensional systems},\ }\href
  {https://doi.org/10.1103/PhysRevX.3.031005} {\bibfield  {journal} {\bibinfo
  {journal} {Phys. Rev. X}\ }\textbf {\bibinfo {volume} {3}},\ \bibinfo {pages}
  {031005} (\bibinfo {year} {2013})}\BibitemShut {NoStop}%
\bibitem [{\citenamefont {Ezawa}(2013)}]{PhysRevLett.110.026603}%
  \BibitemOpen
  \bibfield  {author} {\bibinfo {author} {\bibfnamefont {M.}~\bibnamefont
  {Ezawa}},\ }\bibfield  {title} {\bibinfo {title} {Photoinduced topological
  phase transition and a single {D}irac-cone state in silicene},\ }\href
  {https://doi.org/10.1103/PhysRevLett.110.026603} {\bibfield  {journal}
  {\bibinfo  {journal} {Phys. Rev. Lett.}\ }\textbf {\bibinfo {volume} {110}},\
  \bibinfo {pages} {026603} (\bibinfo {year} {2013})}\BibitemShut {NoStop}%
\bibitem [{\citenamefont {G\'omez-Le\'on}\ and\ \citenamefont
  {Platero}(2013)}]{PhysRevLett.110.200403}%
  \BibitemOpen
  \bibfield  {author} {\bibinfo {author} {\bibfnamefont {A.}~\bibnamefont
  {G\'omez-Le\'on}}\ and\ \bibinfo {author} {\bibfnamefont {G.}~\bibnamefont
  {Platero}},\ }\bibfield  {title} {\bibinfo {title} {Floquet-{B}loch theory
  and topology in periodically driven lattices},\ }\href
  {https://doi.org/10.1103/PhysRevLett.110.200403} {\bibfield  {journal}
  {\bibinfo  {journal} {Phys. Rev. Lett.}\ }\textbf {\bibinfo {volume} {110}},\
  \bibinfo {pages} {200403} (\bibinfo {year} {2013})}\BibitemShut {NoStop}%
\bibitem [{\citenamefont {Grushin}\ \emph {et~al.}(2014)\citenamefont
  {Grushin}, \citenamefont {G\'omez-Le\'on},\ and\ \citenamefont
  {Neupert}}]{PhysRevLett.112.156801}%
  \BibitemOpen
  \bibfield  {author} {\bibinfo {author} {\bibfnamefont {A.~G.}\ \bibnamefont
  {Grushin}}, \bibinfo {author} {\bibfnamefont {A.}~\bibnamefont
  {G\'omez-Le\'on}},\ and\ \bibinfo {author} {\bibfnamefont {T.}~\bibnamefont
  {Neupert}},\ }\bibfield  {title} {\bibinfo {title} {Floquet fractional
  {C}hern insulators},\ }\href {https://doi.org/10.1103/PhysRevLett.112.156801}
  {\bibfield  {journal} {\bibinfo  {journal} {Phys. Rev. Lett.}\ }\textbf
  {\bibinfo {volume} {112}},\ \bibinfo {pages} {156801} (\bibinfo {year}
  {2014})}\BibitemShut {NoStop}%
\bibitem [{\citenamefont {Wang}\ \emph {et~al.}(2014)\citenamefont {Wang},
  \citenamefont {Wang}, \citenamefont {Shen}, \citenamefont {Sheng},\ and\
  \citenamefont {Xing}}]{Wang_2014}%
  \BibitemOpen
  \bibfield  {author} {\bibinfo {author} {\bibfnamefont {R.}~\bibnamefont
  {Wang}}, \bibinfo {author} {\bibfnamefont {B.}~\bibnamefont {Wang}}, \bibinfo
  {author} {\bibfnamefont {R.}~\bibnamefont {Shen}}, \bibinfo {author}
  {\bibfnamefont {L.}~\bibnamefont {Sheng}},\ and\ \bibinfo {author}
  {\bibfnamefont {D.~Y.}\ \bibnamefont {Xing}},\ }\bibfield  {title} {\bibinfo
  {title} {Floquet {W}eyl semimetal induced by off-resonant light},\ }\href
  {https://doi.org/10.1209/0295-5075/105/17004} {\bibfield  {journal} {\bibinfo
   {journal} {{EPL}}\ }\textbf {\bibinfo {volume} {105}},\ \bibinfo {pages}
  {17004} (\bibinfo {year} {2014})}\BibitemShut {NoStop}%
\bibitem [{\citenamefont {Narayan}(2016)}]{PhysRevB.94.041409}%
  \BibitemOpen
  \bibfield  {author} {\bibinfo {author} {\bibfnamefont {A.}~\bibnamefont
  {Narayan}},\ }\bibfield  {title} {\bibinfo {title} {Tunable point nodes from
  line-node semimetals via application of light},\ }\href
  {https://doi.org/10.1103/PhysRevB.94.041409} {\bibfield  {journal} {\bibinfo
  {journal} {Phys. Rev. B}\ }\textbf {\bibinfo {volume} {94}},\ \bibinfo
  {pages} {041409} (\bibinfo {year} {2016})}\BibitemShut {NoStop}%
\bibitem [{\citenamefont {Yan}\ and\ \citenamefont
  {Wang}(2016)}]{PhysRevLett.117.087402}%
  \BibitemOpen
  \bibfield  {author} {\bibinfo {author} {\bibfnamefont {Z.}~\bibnamefont
  {Yan}}\ and\ \bibinfo {author} {\bibfnamefont {Z.}~\bibnamefont {Wang}},\
  }\bibfield  {title} {\bibinfo {title} {Tunable {W}eyl points in periodically
  driven nodal line semimetals},\ }\href
  {https://doi.org/10.1103/PhysRevLett.117.087402} {\bibfield  {journal}
  {\bibinfo  {journal} {Phys. Rev. Lett.}\ }\textbf {\bibinfo {volume} {117}},\
  \bibinfo {pages} {087402} (\bibinfo {year} {2016})}\BibitemShut {NoStop}%
\bibitem [{\citenamefont {Taguchi}\ \emph {et~al.}(2016)\citenamefont
  {Taguchi}, \citenamefont {Xu}, \citenamefont {Yamakage},\ and\ \citenamefont
  {Law}}]{PhysRevB.94.155206}%
  \BibitemOpen
  \bibfield  {author} {\bibinfo {author} {\bibfnamefont {K.}~\bibnamefont
  {Taguchi}}, \bibinfo {author} {\bibfnamefont {D.-H.}\ \bibnamefont {Xu}},
  \bibinfo {author} {\bibfnamefont {A.}~\bibnamefont {Yamakage}},\ and\
  \bibinfo {author} {\bibfnamefont {K.~T.}\ \bibnamefont {Law}},\ }\bibfield
  {title} {\bibinfo {title} {Photovoltaic anomalous {H}all effect in line-node
  semimetals},\ }\href {https://doi.org/10.1103/PhysRevB.94.155206} {\bibfield
  {journal} {\bibinfo  {journal} {Phys. Rev. B}\ }\textbf {\bibinfo {volume}
  {94}},\ \bibinfo {pages} {155206} (\bibinfo {year} {2016})}\BibitemShut
  {NoStop}%
\bibitem [{\citenamefont {Gonz\'alez}\ and\ \citenamefont
  {Molina}(2016)}]{PhysRevLett.116.156803}%
  \BibitemOpen
  \bibfield  {author} {\bibinfo {author} {\bibfnamefont {J.}~\bibnamefont
  {Gonz\'alez}}\ and\ \bibinfo {author} {\bibfnamefont {R.~A.}\ \bibnamefont
  {Molina}},\ }\bibfield  {title} {\bibinfo {title} {Macroscopic degeneracy of
  zero-mode rotating surface states in 3{D} {D}irac and {W}eyl semimetals under
  radiation},\ }\href {https://doi.org/10.1103/PhysRevLett.116.156803}
  {\bibfield  {journal} {\bibinfo  {journal} {Phys. Rev. Lett.}\ }\textbf
  {\bibinfo {volume} {116}},\ \bibinfo {pages} {156803} (\bibinfo {year}
  {2016})}\BibitemShut {NoStop}%
\bibitem [{\citenamefont {Narayan}(2015)}]{PhysRevB.91.205445}%
  \BibitemOpen
  \bibfield  {author} {\bibinfo {author} {\bibfnamefont {A.}~\bibnamefont
  {Narayan}},\ }\bibfield  {title} {\bibinfo {title} {Floquet dynamics in
  two-dimensional semi-{D}irac semimetals and three-dimensional {D}irac
  semimetals},\ }\href {https://doi.org/10.1103/PhysRevB.91.205445} {\bibfield
  {journal} {\bibinfo  {journal} {Phys. Rev. B}\ }\textbf {\bibinfo {volume}
  {91}},\ \bibinfo {pages} {205445} (\bibinfo {year} {2015})}\BibitemShut
  {NoStop}%
\bibitem [{\citenamefont {Saha}(2016)}]{PhysRevB.94.081103}%
  \BibitemOpen
  \bibfield  {author} {\bibinfo {author} {\bibfnamefont {K.}~\bibnamefont
  {Saha}},\ }\bibfield  {title} {\bibinfo {title} {Photoinduced chern
  insulating states in semi-{D}irac materials},\ }\href
  {https://doi.org/10.1103/PhysRevB.94.081103} {\bibfield  {journal} {\bibinfo
  {journal} {Phys. Rev. B}\ }\textbf {\bibinfo {volume} {94}},\ \bibinfo
  {pages} {081103} (\bibinfo {year} {2016})}\BibitemShut {NoStop}%
\bibitem [{\citenamefont {Chan}\ \emph {et~al.}(2016)\citenamefont {Chan},
  \citenamefont {Oh}, \citenamefont {Han},\ and\ \citenamefont
  {Lee}}]{PhysRevB.94.121106}%
  \BibitemOpen
  \bibfield  {author} {\bibinfo {author} {\bibfnamefont {C.-K.}\ \bibnamefont
  {Chan}}, \bibinfo {author} {\bibfnamefont {Y.-T.}\ \bibnamefont {Oh}},
  \bibinfo {author} {\bibfnamefont {J.~H.}\ \bibnamefont {Han}},\ and\ \bibinfo
  {author} {\bibfnamefont {P.~A.}\ \bibnamefont {Lee}},\ }\bibfield  {title}
  {\bibinfo {title} {Type-{II} {W}eyl cone transitions in driven semimetals},\
  }\href {https://doi.org/10.1103/PhysRevB.94.121106} {\bibfield  {journal}
  {\bibinfo  {journal} {Phys. Rev. B}\ }\textbf {\bibinfo {volume} {94}},\
  \bibinfo {pages} {121106} (\bibinfo {year} {2016})}\BibitemShut {NoStop}%
\bibitem [{\citenamefont {Yan}\ and\ \citenamefont
  {Wang}(2017)}]{PhysRevB.96.041206}%
  \BibitemOpen
  \bibfield  {author} {\bibinfo {author} {\bibfnamefont {Z.}~\bibnamefont
  {Yan}}\ and\ \bibinfo {author} {\bibfnamefont {Z.}~\bibnamefont {Wang}},\
  }\bibfield  {title} {\bibinfo {title} {Floquet multi-{W}eyl points in
  crossing-nodal-line semimetals},\ }\href
  {https://doi.org/10.1103/PhysRevB.96.041206} {\bibfield  {journal} {\bibinfo
  {journal} {Phys. Rev. B}\ }\textbf {\bibinfo {volume} {96}},\ \bibinfo
  {pages} {041206} (\bibinfo {year} {2017})}\BibitemShut {NoStop}%
\bibitem [{\citenamefont {Zou}\ and\ \citenamefont
  {Liu}(2016)}]{PhysRevB.93.205435}%
  \BibitemOpen
  \bibfield  {author} {\bibinfo {author} {\bibfnamefont {J.-Y.}\ \bibnamefont
  {Zou}}\ and\ \bibinfo {author} {\bibfnamefont {B.-G.}\ \bibnamefont {Liu}},\
  }\bibfield  {title} {\bibinfo {title} {Floquet {W}eyl fermions in
  three-dimensional stacked graphene systems irradiated by circularly polarized
  light},\ }\href {https://doi.org/10.1103/PhysRevB.93.205435} {\bibfield
  {journal} {\bibinfo  {journal} {Phys. Rev. B}\ }\textbf {\bibinfo {volume}
  {93}},\ \bibinfo {pages} {205435} (\bibinfo {year} {2016})}\BibitemShut
  {NoStop}%
\bibitem [{\citenamefont {H{\"u}bener}\ \emph {et~al.}(2017)\citenamefont
  {H{\"u}bener}, \citenamefont {Sentef}, \citenamefont {De~Giovannini},
  \citenamefont {Kemper},\ and\ \citenamefont {Rubio}}]{hubener2017creating}%
  \BibitemOpen
  \bibfield  {author} {\bibinfo {author} {\bibfnamefont {H.}~\bibnamefont
  {H{\"u}bener}}, \bibinfo {author} {\bibfnamefont {M.~A.}\ \bibnamefont
  {Sentef}}, \bibinfo {author} {\bibfnamefont {U.}~\bibnamefont
  {De~Giovannini}}, \bibinfo {author} {\bibfnamefont {A.~F.}\ \bibnamefont
  {Kemper}},\ and\ \bibinfo {author} {\bibfnamefont {A.}~\bibnamefont
  {Rubio}},\ }\bibfield  {title} {\bibinfo {title} {Creating stable
  {F}loquet--{W}eyl semimetals by laser-driving of 3{D} {D}irac materials},\
  }\href {https://doi.org/10.1038/ncomms13940} {\bibfield  {journal} {\bibinfo
  {journal} {Nat. Commun.}\ }\textbf {\bibinfo {volume} {8}},\ \bibinfo {pages}
  {1} (\bibinfo {year} {2017})}\BibitemShut {NoStop}%
\bibitem [{\citenamefont {Chen}\ \emph {et~al.}(2018)\citenamefont {Chen},
  \citenamefont {Zhou},\ and\ \citenamefont {Xu}}]{PhysRevB.97.155152}%
  \BibitemOpen
  \bibfield  {author} {\bibinfo {author} {\bibfnamefont {R.}~\bibnamefont
  {Chen}}, \bibinfo {author} {\bibfnamefont {B.}~\bibnamefont {Zhou}},\ and\
  \bibinfo {author} {\bibfnamefont {D.-H.}\ \bibnamefont {Xu}},\ }\bibfield
  {title} {\bibinfo {title} {Floquet {W}eyl semimetals in light-irradiated
  type-{II} and hybrid line-node semimetals},\ }\href
  {https://doi.org/10.1103/PhysRevB.97.155152} {\bibfield  {journal} {\bibinfo
  {journal} {Phys. Rev. B}\ }\textbf {\bibinfo {volume} {97}},\ \bibinfo
  {pages} {155152} (\bibinfo {year} {2018})}\BibitemShut {NoStop}%
\bibitem [{\citenamefont {Zhang}\ \emph {et~al.}(2018)\citenamefont {Zhang},
  \citenamefont {Wang}, \citenamefont {Ruan}, \citenamefont {Yao},\ and\
  \citenamefont {Zhang}}]{PhysRevB.97.195139}%
  \BibitemOpen
  \bibfield  {author} {\bibinfo {author} {\bibfnamefont {D.}~\bibnamefont
  {Zhang}}, \bibinfo {author} {\bibfnamefont {H.}~\bibnamefont {Wang}},
  \bibinfo {author} {\bibfnamefont {J.}~\bibnamefont {Ruan}}, \bibinfo {author}
  {\bibfnamefont {G.}~\bibnamefont {Yao}},\ and\ \bibinfo {author}
  {\bibfnamefont {H.}~\bibnamefont {Zhang}},\ }\bibfield  {title} {\bibinfo
  {title} {Engineering topological phases in the {L}uttinger semimetal
  $\ensuremath{\alpha}$-{S}n},\ }\href
  {https://doi.org/10.1103/PhysRevB.97.195139} {\bibfield  {journal} {\bibinfo
  {journal} {Phys. Rev. B}\ }\textbf {\bibinfo {volume} {97}},\ \bibinfo
  {pages} {195139} (\bibinfo {year} {2018})}\BibitemShut {NoStop}%
\bibitem [{\citenamefont {Firoz~Islam}\ and\ \citenamefont
  {Zyuzin}(2019)}]{PhysRevB.100.165302}%
  \BibitemOpen
  \bibfield  {author} {\bibinfo {author} {\bibfnamefont {S.}~\bibnamefont
  {Firoz~Islam}}\ and\ \bibinfo {author} {\bibfnamefont {A.~A.}\ \bibnamefont
  {Zyuzin}},\ }\bibfield  {title} {\bibinfo {title} {Photoinduced interfacial
  chiral modes in threefold topological semimetal},\ }\href
  {https://doi.org/10.1103/PhysRevB.100.165302} {\bibfield  {journal} {\bibinfo
   {journal} {Phys. Rev. B}\ }\textbf {\bibinfo {volume} {100}},\ \bibinfo
  {pages} {165302} (\bibinfo {year} {2019})}\BibitemShut {NoStop}%
\bibitem [{\citenamefont {Deng}\ \emph {et~al.}(2020)\citenamefont {Deng},
  \citenamefont {Zheng}, \citenamefont {Zhan}, \citenamefont {Fan},
  \citenamefont {Wu},\ and\ \citenamefont {Wang}}]{PhysRevB.102.201105}%
  \BibitemOpen
  \bibfield  {author} {\bibinfo {author} {\bibfnamefont {T.}~\bibnamefont
  {Deng}}, \bibinfo {author} {\bibfnamefont {B.}~\bibnamefont {Zheng}},
  \bibinfo {author} {\bibfnamefont {F.}~\bibnamefont {Zhan}}, \bibinfo {author}
  {\bibfnamefont {J.}~\bibnamefont {Fan}}, \bibinfo {author} {\bibfnamefont
  {X.}~\bibnamefont {Wu}},\ and\ \bibinfo {author} {\bibfnamefont
  {R.}~\bibnamefont {Wang}},\ }\bibfield  {title} {\bibinfo {title}
  {Photoinduced {F}loquet mixed-{W}eyl semimetallic phase in a carbon
  allotrope},\ }\href {https://doi.org/10.1103/PhysRevB.102.201105} {\bibfield
  {journal} {\bibinfo  {journal} {Phys. Rev. B}\ }\textbf {\bibinfo {volume}
  {102}},\ \bibinfo {pages} {201105} (\bibinfo {year} {2020})}\BibitemShut
  {NoStop}%
\bibitem [{\citenamefont {Ghosh}\ \emph
  {et~al.}(2022{\natexlab{a}})\citenamefont {Ghosh}, \citenamefont {Nag},\ and\
  \citenamefont {Saha}}]{Ghosh2022systematic}%
  \BibitemOpen
  \bibfield  {author} {\bibinfo {author} {\bibfnamefont {A.~K.}\ \bibnamefont
  {Ghosh}}, \bibinfo {author} {\bibfnamefont {T.}~\bibnamefont {Nag}},\ and\
  \bibinfo {author} {\bibfnamefont {A.}~\bibnamefont {Saha}},\ }\bibfield
  {title} {\bibinfo {title} {Systematic generation of the cascade of anomalous
  dynamical first- and higher-order modes in floquet topological insulators},\
  }\href {https://doi.org/10.1103/PhysRevB.105.115418} {\bibfield  {journal}
  {\bibinfo  {journal} {Phys. Rev. B}\ }\textbf {\bibinfo {volume} {105}},\
  \bibinfo {pages} {115418} (\bibinfo {year} {2022}{\natexlab{a}})}\BibitemShut
  {NoStop}%
\bibitem [{\citenamefont {Nag}\ and\ \citenamefont
  {Roy}(2021)}]{nag2021anomalous}%
  \BibitemOpen
  \bibfield  {author} {\bibinfo {author} {\bibfnamefont {T.}~\bibnamefont
  {Nag}}\ and\ \bibinfo {author} {\bibfnamefont {B.}~\bibnamefont {Roy}},\
  }\bibfield  {title} {\bibinfo {title} {Anomalous and normal dislocation modes
  in floquet topological insulators},\ }\href
  {https://doi.org/10.1038/s42005-021-00659-4} {\bibfield  {journal} {\bibinfo
  {journal} {Commun. Phys.}\ }\textbf {\bibinfo {volume} {4}},\ \bibinfo
  {pages} {1} (\bibinfo {year} {2021})}\BibitemShut {NoStop}%
\bibitem [{\citenamefont {Nag}\ \emph {et~al.}(2019{\natexlab{a}})\citenamefont
  {Nag}, \citenamefont {Slager}, \citenamefont {Higuchi},\ and\ \citenamefont
  {Oka}}]{Nag2019dynamical}%
  \BibitemOpen
  \bibfield  {author} {\bibinfo {author} {\bibfnamefont {T.}~\bibnamefont
  {Nag}}, \bibinfo {author} {\bibfnamefont {R.-J.}\ \bibnamefont {Slager}},
  \bibinfo {author} {\bibfnamefont {T.}~\bibnamefont {Higuchi}},\ and\ \bibinfo
  {author} {\bibfnamefont {T.}~\bibnamefont {Oka}},\ }\bibfield  {title}
  {\bibinfo {title} {Dynamical synchronization transition in interacting
  electron systems},\ }\href {https://doi.org/10.1103/PhysRevB.100.134301}
  {\bibfield  {journal} {\bibinfo  {journal} {Phys. Rev. B}\ }\textbf {\bibinfo
  {volume} {100}},\ \bibinfo {pages} {134301} (\bibinfo {year}
  {2019}{\natexlab{a}})}\BibitemShut {NoStop}%
\bibitem [{\citenamefont {Du}\ \emph {et~al.}(2022)\citenamefont {Du},
  \citenamefont {Chen}, \citenamefont {Wang},\ and\ \citenamefont
  {Xu}}]{XLDU2022}%
  \BibitemOpen
  \bibfield  {author} {\bibinfo {author} {\bibfnamefont {X.-L.}\ \bibnamefont
  {Du}}, \bibinfo {author} {\bibfnamefont {R.}~\bibnamefont {Chen}}, \bibinfo
  {author} {\bibfnamefont {R.}~\bibnamefont {Wang}},\ and\ \bibinfo {author}
  {\bibfnamefont {D.-H.}\ \bibnamefont {Xu}},\ }\bibfield  {title} {\bibinfo
  {title} {Weyl nodes with higher-order topology in an optically driven
  nodal-line semimetal},\ }\href {https://doi.org/10.1103/PhysRevB.105.L081102}
  {\bibfield  {journal} {\bibinfo  {journal} {Phys. Rev. B}\ }\textbf {\bibinfo
  {volume} {105}},\ \bibinfo {pages} {L081102} (\bibinfo {year}
  {2022})}\BibitemShut {NoStop}%
\bibitem [{\citenamefont {Wang}\ \emph {et~al.}(2023)\citenamefont {Wang},
  \citenamefont {Wang}, \citenamefont {Sun}, \citenamefont {Chen},\ and\
  \citenamefont {Xu}}]{ZMWangFloquet}%
  \BibitemOpen
  \bibfield  {author} {\bibinfo {author} {\bibfnamefont {Z.-M.}\ \bibnamefont
  {Wang}}, \bibinfo {author} {\bibfnamefont {R.}~\bibnamefont {Wang}}, \bibinfo
  {author} {\bibfnamefont {J.-H.}\ \bibnamefont {Sun}}, \bibinfo {author}
  {\bibfnamefont {T.-Y.}\ \bibnamefont {Chen}},\ and\ \bibinfo {author}
  {\bibfnamefont {D.-H.}\ \bibnamefont {Xu}},\ }\bibfield  {title} {\bibinfo
  {title} {Floquet weyl semimetal phases in light-irradiated higher-order
  topological dirac semimetals},\ }\href
  {https://doi.org/10.1103/PhysRevB.107.L121407} {\bibfield  {journal}
  {\bibinfo  {journal} {Phys. Rev. B}\ }\textbf {\bibinfo {volume} {107}},\
  \bibinfo {pages} {L121407} (\bibinfo {year} {2023})}\BibitemShut {NoStop}%
\bibitem [{\citenamefont {Zhan}\ \emph {et~al.}(2024)\citenamefont {Zhan},
  \citenamefont {Chen}, \citenamefont {Ning}, \citenamefont {Ma}, \citenamefont
  {Wang}, \citenamefont {Xu},\ and\ \citenamefont
  {Wang}}]{zhan2024perspective}%
  \BibitemOpen
  \bibfield  {author} {\bibinfo {author} {\bibfnamefont {F.}~\bibnamefont
  {Zhan}}, \bibinfo {author} {\bibfnamefont {R.}~\bibnamefont {Chen}}, \bibinfo
  {author} {\bibfnamefont {Z.}~\bibnamefont {Ning}}, \bibinfo {author}
  {\bibfnamefont {D.-S.}\ \bibnamefont {Ma}}, \bibinfo {author} {\bibfnamefont
  {Z.}~\bibnamefont {Wang}}, \bibinfo {author} {\bibfnamefont {D.-H.}\
  \bibnamefont {Xu}},\ and\ \bibinfo {author} {\bibfnamefont {R.}~\bibnamefont
  {Wang}},\ }\bibfield  {title} {\bibinfo {title} {Perspective: Floquet
  engineering topological states from effective models towards realistic
  materials},\ }\href {https://doi.org/10.1007/s44214-024-00067-z} {\bibfield
  {journal} {\bibinfo  {journal} {Quantum Front.}\ }\textbf {\bibinfo {volume}
  {3}},\ \bibinfo {pages} {21} (\bibinfo {year} {2024})}\BibitemShut {NoStop}%
\bibitem [{\citenamefont {Ning}\ \emph {et~al.}(2024)\citenamefont {Ning},
  \citenamefont {Ma}, \citenamefont {Zeng}, \citenamefont {Xu},\ and\
  \citenamefont {Wang}}]{Zengprl}%
  \BibitemOpen
  \bibfield  {author} {\bibinfo {author} {\bibfnamefont {Z.}~\bibnamefont
  {Ning}}, \bibinfo {author} {\bibfnamefont {D.-S.}\ \bibnamefont {Ma}},
  \bibinfo {author} {\bibfnamefont {J.}~\bibnamefont {Zeng}}, \bibinfo {author}
  {\bibfnamefont {D.-H.}\ \bibnamefont {Xu}},\ and\ \bibinfo {author}
  {\bibfnamefont {R.}~\bibnamefont {Wang}},\ }\bibfield  {title} {\bibinfo
  {title} {Flexible control of chiral superconductivity in optically driven
  nodal point superconductors with antiferromagnetism},\ }\href
  {https://doi.org/10.1103/PhysRevLett.133.246606} {\bibfield  {journal}
  {\bibinfo  {journal} {Phys. Rev. Lett.}\ }\textbf {\bibinfo {volume} {133}},\
  \bibinfo {pages} {246606} (\bibinfo {year} {2024})}\BibitemShut {NoStop}%
\bibitem [{\citenamefont {Trevisan}\ \emph {et~al.}(2022)\citenamefont
  {Trevisan}, \citenamefont {Arribi}, \citenamefont {Heinonen}, \citenamefont
  {Slager},\ and\ \citenamefont {Orth}}]{PhysRevLett.128.066602}%
  \BibitemOpen
  \bibfield  {author} {\bibinfo {author} {\bibfnamefont {T.~V.}\ \bibnamefont
  {Trevisan}}, \bibinfo {author} {\bibfnamefont {P.~V.}\ \bibnamefont
  {Arribi}}, \bibinfo {author} {\bibfnamefont {O.}~\bibnamefont {Heinonen}},
  \bibinfo {author} {\bibfnamefont {R.-J.}\ \bibnamefont {Slager}},\ and\
  \bibinfo {author} {\bibfnamefont {P.~P.}\ \bibnamefont {Orth}},\ }\bibfield
  {title} {\bibinfo {title} {Bicircular light {F}loquet engineering of magnetic
  symmetry and topology and its application to the {D}irac semimetal
  {C}d$_{3}${A}s$_{2}$},\ }\href
  {https://doi.org/10.1103/PhysRevLett.128.066602} {\bibfield  {journal}
  {\bibinfo  {journal} {Phys. Rev. Lett.}\ }\textbf {\bibinfo {volume} {128}},\
  \bibinfo {pages} {066602} (\bibinfo {year} {2022})}\BibitemShut {NoStop}%
\bibitem [{\citenamefont {Bomantara}\ \emph {et~al.}(2019)\citenamefont
  {Bomantara}, \citenamefont {Zhou}, \citenamefont {Pan},\ and\ \citenamefont
  {Gong}}]{PhysRevB.99.045441}%
  \BibitemOpen
  \bibfield  {author} {\bibinfo {author} {\bibfnamefont {R.~W.}\ \bibnamefont
  {Bomantara}}, \bibinfo {author} {\bibfnamefont {L.}~\bibnamefont {Zhou}},
  \bibinfo {author} {\bibfnamefont {J.}~\bibnamefont {Pan}},\ and\ \bibinfo
  {author} {\bibfnamefont {J.}~\bibnamefont {Gong}},\ }\bibfield  {title}
  {\bibinfo {title} {Coupled-wire construction of static and {F}loquet
  second-order topological insulators},\ }\href
  {https://doi.org/10.1103/PhysRevB.99.045441} {\bibfield  {journal} {\bibinfo
  {journal} {Phys. Rev. B}\ }\textbf {\bibinfo {volume} {99}},\ \bibinfo
  {pages} {045441} (\bibinfo {year} {2019})}\BibitemShut {NoStop}%
\bibitem [{\citenamefont {Rodriguez-Vega}\ \emph {et~al.}(2019)\citenamefont
  {Rodriguez-Vega}, \citenamefont {Kumar},\ and\ \citenamefont
  {Seradjeh}}]{PhysRevB.100.085138}%
  \BibitemOpen
  \bibfield  {author} {\bibinfo {author} {\bibfnamefont {M.}~\bibnamefont
  {Rodriguez-Vega}}, \bibinfo {author} {\bibfnamefont {A.}~\bibnamefont
  {Kumar}},\ and\ \bibinfo {author} {\bibfnamefont {B.}~\bibnamefont
  {Seradjeh}},\ }\bibfield  {title} {\bibinfo {title} {Higher-order {F}loquet
  topological phases with corner and bulk bound states},\ }\href
  {https://doi.org/10.1103/PhysRevB.100.085138} {\bibfield  {journal} {\bibinfo
   {journal} {Phys. Rev. B}\ }\textbf {\bibinfo {volume} {100}},\ \bibinfo
  {pages} {085138} (\bibinfo {year} {2019})}\BibitemShut {NoStop}%
\bibitem [{\citenamefont {Nag}\ \emph {et~al.}(2019{\natexlab{b}})\citenamefont
  {Nag}, \citenamefont {Juri\ifmmode \check{c}\else
  \v{c}\fi{}i\ifmmode~\acute{c}\else \'{c}\fi{}},\ and\ \citenamefont
  {Roy}}]{PhysRevResearch.1.032045}%
  \BibitemOpen
  \bibfield  {author} {\bibinfo {author} {\bibfnamefont {T.}~\bibnamefont
  {Nag}}, \bibinfo {author} {\bibfnamefont {V.}~\bibnamefont {Juri\ifmmode
  \check{c}\else \v{c}\fi{}i\ifmmode~\acute{c}\else \'{c}\fi{}}},\ and\
  \bibinfo {author} {\bibfnamefont {B.}~\bibnamefont {Roy}},\ }\bibfield
  {title} {\bibinfo {title} {Out of equilibrium higher-order topological
  insulator: {F}loquet engineering and quench dynamics},\ }\href
  {https://doi.org/10.1103/PhysRevResearch.1.032045} {\bibfield  {journal}
  {\bibinfo  {journal} {Phys. Rev. Research}\ }\textbf {\bibinfo {volume}
  {1}},\ \bibinfo {pages} {032045} (\bibinfo {year}
  {2019}{\natexlab{b}})}\BibitemShut {NoStop}%
\bibitem [{\citenamefont {Seshadri}\ \emph {et~al.}(2019)\citenamefont
  {Seshadri}, \citenamefont {Dutta},\ and\ \citenamefont
  {Sen}}]{PhysRevB.100.115403}%
  \BibitemOpen
  \bibfield  {author} {\bibinfo {author} {\bibfnamefont {R.}~\bibnamefont
  {Seshadri}}, \bibinfo {author} {\bibfnamefont {A.}~\bibnamefont {Dutta}},\
  and\ \bibinfo {author} {\bibfnamefont {D.}~\bibnamefont {Sen}},\ }\bibfield
  {title} {\bibinfo {title} {Generating a second-order topological insulator
  with multiple corner states by periodic driving},\ }\href
  {https://doi.org/10.1103/PhysRevB.100.115403} {\bibfield  {journal} {\bibinfo
   {journal} {Phys. Rev. B}\ }\textbf {\bibinfo {volume} {100}},\ \bibinfo
  {pages} {115403} (\bibinfo {year} {2019})}\BibitemShut {NoStop}%
\bibitem [{\citenamefont {Peng}\ and\ \citenamefont
  {Refael}(2019)}]{PhysRevLett.123.016806}%
  \BibitemOpen
  \bibfield  {author} {\bibinfo {author} {\bibfnamefont {Y.}~\bibnamefont
  {Peng}}\ and\ \bibinfo {author} {\bibfnamefont {G.}~\bibnamefont {Refael}},\
  }\bibfield  {title} {\bibinfo {title} {Floquet second-order topological
  insulators from nonsymmorphic space-time symmetries},\ }\href
  {https://doi.org/10.1103/PhysRevLett.123.016806} {\bibfield  {journal}
  {\bibinfo  {journal} {Phys. Rev. Lett.}\ }\textbf {\bibinfo {volume} {123}},\
  \bibinfo {pages} {016806} (\bibinfo {year} {2019})}\BibitemShut {NoStop}%
\bibitem [{\citenamefont {Hu}\ \emph {et~al.}(2020)\citenamefont {Hu},
  \citenamefont {Huang}, \citenamefont {Zhao},\ and\ \citenamefont
  {Liu}}]{PhysRevLett.124.057001}%
  \BibitemOpen
  \bibfield  {author} {\bibinfo {author} {\bibfnamefont {H.}~\bibnamefont
  {Hu}}, \bibinfo {author} {\bibfnamefont {B.}~\bibnamefont {Huang}}, \bibinfo
  {author} {\bibfnamefont {E.}~\bibnamefont {Zhao}},\ and\ \bibinfo {author}
  {\bibfnamefont {W.~V.}\ \bibnamefont {Liu}},\ }\bibfield  {title} {\bibinfo
  {title} {Dynamical singularities of {F}loquet higher-order topological
  insulators},\ }\href {https://doi.org/10.1103/PhysRevLett.124.057001}
  {\bibfield  {journal} {\bibinfo  {journal} {Phys. Rev. Lett.}\ }\textbf
  {\bibinfo {volume} {124}},\ \bibinfo {pages} {057001} (\bibinfo {year}
  {2020})}\BibitemShut {NoStop}%
\bibitem [{\citenamefont {Ghosh}\ \emph {et~al.}(2020)\citenamefont {Ghosh},
  \citenamefont {Paul},\ and\ \citenamefont {Saha}}]{PhysRevB.101.235403}%
  \BibitemOpen
  \bibfield  {author} {\bibinfo {author} {\bibfnamefont {A.~K.}\ \bibnamefont
  {Ghosh}}, \bibinfo {author} {\bibfnamefont {G.~C.}\ \bibnamefont {Paul}},\
  and\ \bibinfo {author} {\bibfnamefont {A.}~\bibnamefont {Saha}},\ }\bibfield
  {title} {\bibinfo {title} {Higher order topological insulator via periodic
  driving},\ }\href {https://doi.org/10.1103/PhysRevB.101.235403} {\bibfield
  {journal} {\bibinfo  {journal} {Phys. Rev. B}\ }\textbf {\bibinfo {volume}
  {101}},\ \bibinfo {pages} {235403} (\bibinfo {year} {2020})}\BibitemShut
  {NoStop}%
\bibitem [{\citenamefont {Peng}(2020)}]{PhysRevResearch.2.013124}%
  \BibitemOpen
  \bibfield  {author} {\bibinfo {author} {\bibfnamefont {Y.}~\bibnamefont
  {Peng}},\ }\bibfield  {title} {\bibinfo {title} {Floquet higher-order
  topological insulators and superconductors with space-time symmetries},\
  }\href {https://doi.org/10.1103/PhysRevResearch.2.013124} {\bibfield
  {journal} {\bibinfo  {journal} {Phys. Rev. Research}\ }\textbf {\bibinfo
  {volume} {2}},\ \bibinfo {pages} {013124} (\bibinfo {year}
  {2020})}\BibitemShut {NoStop}%
\bibitem [{\citenamefont {Bomantara}(2020)}]{PhysRevResearch.2.033495}%
  \BibitemOpen
  \bibfield  {author} {\bibinfo {author} {\bibfnamefont {R.~W.}\ \bibnamefont
  {Bomantara}},\ }\bibfield  {title} {\bibinfo {title} {Time-induced
  second-order topological superconductors},\ }\href
  {https://doi.org/10.1103/PhysRevResearch.2.033495} {\bibfield  {journal}
  {\bibinfo  {journal} {Phys. Rev. Research}\ }\textbf {\bibinfo {volume}
  {2}},\ \bibinfo {pages} {033495} (\bibinfo {year} {2020})}\BibitemShut
  {NoStop}%
\bibitem [{\citenamefont {Huang}\ and\ \citenamefont
  {Liu}(2020)}]{PhysRevLett.124.216601}%
  \BibitemOpen
  \bibfield  {author} {\bibinfo {author} {\bibfnamefont {B.}~\bibnamefont
  {Huang}}\ and\ \bibinfo {author} {\bibfnamefont {W.~V.}\ \bibnamefont
  {Liu}},\ }\bibfield  {title} {\bibinfo {title} {Floquet higher-order
  topological insulators with anomalous dynamical polarization},\ }\href
  {https://doi.org/10.1103/PhysRevLett.124.216601} {\bibfield  {journal}
  {\bibinfo  {journal} {Phys. Rev. Lett.}\ }\textbf {\bibinfo {volume} {124}},\
  \bibinfo {pages} {216601} (\bibinfo {year} {2020})}\BibitemShut {NoStop}%
\bibitem [{\citenamefont {Zhu}\ \emph {et~al.}(2021{\natexlab{a}})\citenamefont
  {Zhu}, \citenamefont {Chong},\ and\ \citenamefont
  {Gong}}]{PhysRevB.103.L041402}%
  \BibitemOpen
  \bibfield  {author} {\bibinfo {author} {\bibfnamefont {W.}~\bibnamefont
  {Zhu}}, \bibinfo {author} {\bibfnamefont {Y.~D.}\ \bibnamefont {Chong}},\
  and\ \bibinfo {author} {\bibfnamefont {J.}~\bibnamefont {Gong}},\ }\bibfield
  {title} {\bibinfo {title} {Floquet higher-order topological insulator in a
  periodically driven bipartite lattice},\ }\href
  {https://doi.org/10.1103/PhysRevB.103.L041402} {\bibfield  {journal}
  {\bibinfo  {journal} {Phys. Rev. B}\ }\textbf {\bibinfo {volume} {103}},\
  \bibinfo {pages} {L041402} (\bibinfo {year}
  {2021}{\natexlab{a}})}\BibitemShut {NoStop}%
\bibitem [{\citenamefont {Zhang}\ and\ \citenamefont
  {Yang}(2021)}]{PhysRevB.103.L121115}%
  \BibitemOpen
  \bibfield  {author} {\bibinfo {author} {\bibfnamefont {R.-X.}\ \bibnamefont
  {Zhang}}\ and\ \bibinfo {author} {\bibfnamefont {Z.-C.}\ \bibnamefont
  {Yang}},\ }\bibfield  {title} {\bibinfo {title} {Tunable fragile topology in
  {F}loquet systems},\ }\href {https://doi.org/10.1103/PhysRevB.103.L121115}
  {\bibfield  {journal} {\bibinfo  {journal} {Phys. Rev. B}\ }\textbf {\bibinfo
  {volume} {103}},\ \bibinfo {pages} {L121115} (\bibinfo {year}
  {2021})}\BibitemShut {NoStop}%
\bibitem [{\citenamefont {Nag}\ \emph {et~al.}(2021)\citenamefont {Nag},
  \citenamefont {Juri\ifmmode \check{c}\else \v{c}\fi{}i\ifmmode~\acute{c}\else
  \'{c}\fi{}},\ and\ \citenamefont {Roy}}]{PhysRevB.103.115308}%
  \BibitemOpen
  \bibfield  {author} {\bibinfo {author} {\bibfnamefont {T.}~\bibnamefont
  {Nag}}, \bibinfo {author} {\bibfnamefont {V.}~\bibnamefont {Juri\ifmmode
  \check{c}\else \v{c}\fi{}i\ifmmode~\acute{c}\else \'{c}\fi{}}},\ and\
  \bibinfo {author} {\bibfnamefont {B.}~\bibnamefont {Roy}},\ }\bibfield
  {title} {\bibinfo {title} {Hierarchy of higher-order {F}loquet topological
  phases in three dimensions},\ }\href
  {https://doi.org/10.1103/PhysRevB.103.115308} {\bibfield  {journal} {\bibinfo
   {journal} {Phys. Rev. B}\ }\textbf {\bibinfo {volume} {103}},\ \bibinfo
  {pages} {115308} (\bibinfo {year} {2021})}\BibitemShut {NoStop}%
\bibitem [{\citenamefont {Zhu}\ \emph {et~al.}(2021{\natexlab{b}})\citenamefont
  {Zhu}, \citenamefont {Chong},\ and\ \citenamefont
  {Gong}}]{PhysRevB.104.L020302}%
  \BibitemOpen
  \bibfield  {author} {\bibinfo {author} {\bibfnamefont {W.}~\bibnamefont
  {Zhu}}, \bibinfo {author} {\bibfnamefont {Y.~D.}\ \bibnamefont {Chong}},\
  and\ \bibinfo {author} {\bibfnamefont {J.}~\bibnamefont {Gong}},\ }\bibfield
  {title} {\bibinfo {title} {Symmetry analysis of anomalous {F}loquet
  topological phases},\ }\href {https://doi.org/10.1103/PhysRevB.104.L020302}
  {\bibfield  {journal} {\bibinfo  {journal} {Phys. Rev. B}\ }\textbf {\bibinfo
  {volume} {104}},\ \bibinfo {pages} {L020302} (\bibinfo {year}
  {2021}{\natexlab{b}})}\BibitemShut {NoStop}%
\bibitem [{\citenamefont {Zhu}\ \emph {et~al.}(2021{\natexlab{c}})\citenamefont
  {Zhu}, \citenamefont {Umer},\ and\ \citenamefont
  {Gong}}]{PhysRevResearch.3.L032026}%
  \BibitemOpen
  \bibfield  {author} {\bibinfo {author} {\bibfnamefont {W.}~\bibnamefont
  {Zhu}}, \bibinfo {author} {\bibfnamefont {M.}~\bibnamefont {Umer}},\ and\
  \bibinfo {author} {\bibfnamefont {J.}~\bibnamefont {Gong}},\ }\bibfield
  {title} {\bibinfo {title} {Floquet higher-order {W}eyl and nexus
  semimetals},\ }\href {https://doi.org/10.1103/PhysRevResearch.3.L032026}
  {\bibfield  {journal} {\bibinfo  {journal} {Phys. Rev. Research}\ }\textbf
  {\bibinfo {volume} {3}},\ \bibinfo {pages} {L032026} (\bibinfo {year}
  {2021}{\natexlab{c}})}\BibitemShut {NoStop}%
\bibitem [{\citenamefont {Rui}\ \emph {et~al.}(2021)\citenamefont {Rui},
  \citenamefont {Zhang}, \citenamefont {Hirschmann}, \citenamefont {Zheng},
  \citenamefont {Schnyder}, \citenamefont {Trauzettel},\ and\ \citenamefont
  {Wang}}]{PhysRevB.103.184510}%
  \BibitemOpen
  \bibfield  {author} {\bibinfo {author} {\bibfnamefont {W.~B.}\ \bibnamefont
  {Rui}}, \bibinfo {author} {\bibfnamefont {S.-B.}\ \bibnamefont {Zhang}},
  \bibinfo {author} {\bibfnamefont {M.~M.}\ \bibnamefont {Hirschmann}},
  \bibinfo {author} {\bibfnamefont {Z.}~\bibnamefont {Zheng}}, \bibinfo
  {author} {\bibfnamefont {A.~P.}\ \bibnamefont {Schnyder}}, \bibinfo {author}
  {\bibfnamefont {B.}~\bibnamefont {Trauzettel}},\ and\ \bibinfo {author}
  {\bibfnamefont {Z.~D.}\ \bibnamefont {Wang}},\ }\bibfield  {title} {\bibinfo
  {title} {Higher-order {W}eyl superconductors with anisotropic {W}eyl-point
  connectivity},\ }\href {https://doi.org/10.1103/PhysRevB.103.184510}
  {\bibfield  {journal} {\bibinfo  {journal} {Phys. Rev. B}\ }\textbf {\bibinfo
  {volume} {103}},\ \bibinfo {pages} {184510} (\bibinfo {year}
  {2021})}\BibitemShut {NoStop}%
\bibitem [{\citenamefont {Wang}\ \emph {et~al.}(2021)\citenamefont {Wang},
  \citenamefont {Wu},\ and\ \citenamefont {An}}]{PhysRevB.104.205117}%
  \BibitemOpen
  \bibfield  {author} {\bibinfo {author} {\bibfnamefont {B.-Q.}\ \bibnamefont
  {Wang}}, \bibinfo {author} {\bibfnamefont {H.}~\bibnamefont {Wu}},\ and\
  \bibinfo {author} {\bibfnamefont {J.-H.}\ \bibnamefont {An}},\ }\bibfield
  {title} {\bibinfo {title} {Engineering exotic second-order topological
  semimetals by periodic driving},\ }\href
  {https://doi.org/10.1103/PhysRevB.104.205117} {\bibfield  {journal} {\bibinfo
   {journal} {Phys. Rev. B}\ }\textbf {\bibinfo {volume} {104}},\ \bibinfo
  {pages} {205117} (\bibinfo {year} {2021})}\BibitemShut {NoStop}%
\bibitem [{\citenamefont {Ghosh}\ \emph
  {et~al.}(2022{\natexlab{b}})\citenamefont {Ghosh}, \citenamefont {Saha},\
  and\ \citenamefont {Sengupta}}]{Ghosh2022hinge-mode}%
  \BibitemOpen
  \bibfield  {author} {\bibinfo {author} {\bibfnamefont {S.}~\bibnamefont
  {Ghosh}}, \bibinfo {author} {\bibfnamefont {K.}~\bibnamefont {Saha}},\ and\
  \bibinfo {author} {\bibfnamefont {K.}~\bibnamefont {Sengupta}},\ }\bibfield
  {title} {\bibinfo {title} {Hinge-mode dynamics of periodically driven
  higher-order weyl semimetals},\ }\href
  {https://doi.org/10.1103/PhysRevB.105.224312} {\bibfield  {journal} {\bibinfo
   {journal} {Phys. Rev. B}\ }\textbf {\bibinfo {volume} {105}},\ \bibinfo
  {pages} {224312} (\bibinfo {year} {2022}{\natexlab{b}})}\BibitemShut
  {NoStop}%
\bibitem [{\citenamefont {Liu}\ \emph {et~al.}(2025)\citenamefont {Liu},
  \citenamefont {Cui}, \citenamefont {Li}, \citenamefont {Li}, \citenamefont
  {Xu},\ and\ \citenamefont {Yu}}]{Liupprb}%
  \BibitemOpen
  \bibfield  {author} {\bibinfo {author} {\bibfnamefont {P.}~\bibnamefont
  {Liu}}, \bibinfo {author} {\bibfnamefont {C.}~\bibnamefont {Cui}}, \bibinfo
  {author} {\bibfnamefont {L.}~\bibnamefont {Li}}, \bibinfo {author}
  {\bibfnamefont {R.}~\bibnamefont {Li}}, \bibinfo {author} {\bibfnamefont
  {D.-H.}\ \bibnamefont {Xu}},\ and\ \bibinfo {author} {\bibfnamefont {Z.-M.}\
  \bibnamefont {Yu}},\ }\bibfield  {title} {\bibinfo {title} {Floquet control
  of topological phases and hall effects in ${Z}_{2}$ nodal line semimetals},\
  }\href {https://doi.org/10.1103/PhysRevB.111.235105} {\bibfield  {journal}
  {\bibinfo  {journal} {Phys. Rev. B}\ }\textbf {\bibinfo {volume} {111}},\
  \bibinfo {pages} {235105} (\bibinfo {year} {2025})}\BibitemShut {NoStop}%
\bibitem [{\citenamefont {Wan}\ \emph {et~al.}(2026)\citenamefont {Wan},
  \citenamefont {Zhan}, \citenamefont {Ding}, \citenamefont {Qin},
  \citenamefont {Huang}, \citenamefont {Xu},\ and\ \citenamefont
  {Wang}}]{wanprb1}%
  \BibitemOpen
  \bibfield  {author} {\bibinfo {author} {\bibfnamefont {X.}~\bibnamefont
  {Wan}}, \bibinfo {author} {\bibfnamefont {F.}~\bibnamefont {Zhan}}, \bibinfo
  {author} {\bibfnamefont {X.}~\bibnamefont {Ding}}, \bibinfo {author}
  {\bibfnamefont {Z.}~\bibnamefont {Qin}}, \bibinfo {author} {\bibfnamefont
  {S.}~\bibnamefont {Huang}}, \bibinfo {author} {\bibfnamefont {D.-H.}\
  \bibnamefont {Xu}},\ and\ \bibinfo {author} {\bibfnamefont {R.}~\bibnamefont
  {Wang}},\ }\bibfield  {title} {\bibinfo {title} {Floquet higher-order weyl
  semimetallic phase in three-dimensional graphdiyne},\ }\href
  {https://doi.org/10.1103/rp1t-nzsh} {\bibfield  {journal} {\bibinfo
  {journal} {Phys. Rev. B}\ }\textbf {\bibinfo {volume} {113}},\ \bibinfo
  {pages} {L041109} (\bibinfo {year} {2026})}\BibitemShut {NoStop}%
\bibitem [{\citenamefont {Huang}\ \emph {et~al.}(2024)\citenamefont {Huang},
  \citenamefont {Zhan}, \citenamefont {Ding}, \citenamefont {Xu}, \citenamefont
  {Ma},\ and\ \citenamefont {Wang}}]{Huangprb}%
  \BibitemOpen
  \bibfield  {author} {\bibinfo {author} {\bibfnamefont {S.}~\bibnamefont
  {Huang}}, \bibinfo {author} {\bibfnamefont {F.}~\bibnamefont {Zhan}},
  \bibinfo {author} {\bibfnamefont {X.}~\bibnamefont {Ding}}, \bibinfo {author}
  {\bibfnamefont {D.-H.}\ \bibnamefont {Xu}}, \bibinfo {author} {\bibfnamefont
  {D.-S.}\ \bibnamefont {Ma}},\ and\ \bibinfo {author} {\bibfnamefont
  {R.}~\bibnamefont {Wang}},\ }\bibfield  {title} {\bibinfo {title}
  {Controllable weyl nodes and fermi arcs from floquet engineering triple
  fermions},\ }\href {https://doi.org/10.1103/PhysRevB.110.L121118} {\bibfield
  {journal} {\bibinfo  {journal} {Phys. Rev. B}\ }\textbf {\bibinfo {volume}
  {110}},\ \bibinfo {pages} {L121118} (\bibinfo {year} {2024})}\BibitemShut
  {NoStop}%
\bibitem [{\citenamefont {McIver}\ \emph {et~al.}(2020)\citenamefont {McIver},
  \citenamefont {Schulte}, \citenamefont {Stein}, \citenamefont {Matsuyama},
  \citenamefont {Jotzu}, \citenamefont {Meier},\ and\ \citenamefont
  {Cavalleri}}]{mciver2020light}%
  \BibitemOpen
  \bibfield  {author} {\bibinfo {author} {\bibfnamefont {J.~W.}\ \bibnamefont
  {McIver}}, \bibinfo {author} {\bibfnamefont {B.}~\bibnamefont {Schulte}},
  \bibinfo {author} {\bibfnamefont {F.-U.}\ \bibnamefont {Stein}}, \bibinfo
  {author} {\bibfnamefont {T.}~\bibnamefont {Matsuyama}}, \bibinfo {author}
  {\bibfnamefont {G.}~\bibnamefont {Jotzu}}, \bibinfo {author} {\bibfnamefont
  {G.}~\bibnamefont {Meier}},\ and\ \bibinfo {author} {\bibfnamefont
  {A.}~\bibnamefont {Cavalleri}},\ }\bibfield  {title} {\bibinfo {title}
  {Light-induced anomalous {H}all effect in graphene},\ }\href
  {https://doi.org/10.1038/s41567-019-0698-y} {\bibfield  {journal} {\bibinfo
  {journal} {Nat. Phys.}\ }\textbf {\bibinfo {volume} {16}},\ \bibinfo {pages}
  {38} (\bibinfo {year} {2020})}\BibitemShut {NoStop}%
\bibitem [{\citenamefont {Choi}\ \emph {et~al.}(2025)\citenamefont {Choi},
  \citenamefont {Mogi}, \citenamefont {De~Giovannini}, \citenamefont {Azoury},
  \citenamefont {Lv}, \citenamefont {Su}, \citenamefont {H{\"u}bener},
  \citenamefont {Rubio},\ and\ \citenamefont {Gedik}}]{Choi2025}%
  \BibitemOpen
  \bibfield  {author} {\bibinfo {author} {\bibfnamefont {D.}~\bibnamefont
  {Choi}}, \bibinfo {author} {\bibfnamefont {M.}~\bibnamefont {Mogi}}, \bibinfo
  {author} {\bibfnamefont {U.}~\bibnamefont {De~Giovannini}}, \bibinfo {author}
  {\bibfnamefont {D.}~\bibnamefont {Azoury}}, \bibinfo {author} {\bibfnamefont
  {B.}~\bibnamefont {Lv}}, \bibinfo {author} {\bibfnamefont {Y.}~\bibnamefont
  {Su}}, \bibinfo {author} {\bibfnamefont {H.}~\bibnamefont {H{\"u}bener}},
  \bibinfo {author} {\bibfnamefont {A.}~\bibnamefont {Rubio}},\ and\ \bibinfo
  {author} {\bibfnamefont {N.}~\bibnamefont {Gedik}},\ }\bibfield  {title}
  {\bibinfo {title} {Observation of floquet--bloch states in monolayer
  graphene},\ }\href {https://doi.org/10.1038/s41567-025-02888-8} {\bibfield
  {journal} {\bibinfo  {journal} {Nat. Phys.}\ }\textbf {\bibinfo {volume}
  {21}},\ \bibinfo {pages} {1100} (\bibinfo {year} {2025})}\BibitemShut
  {NoStop}%
\bibitem [{\citenamefont {Merboldt}\ \emph {et~al.}(2025)\citenamefont
  {Merboldt}, \citenamefont {Sch{\"u}ler}, \citenamefont {Schmitt},
  \citenamefont {Bange}, \citenamefont {Bennecke}, \citenamefont {Gadge},
  \citenamefont {Pierz}, \citenamefont {Schumacher}, \citenamefont {Momeni},
  \citenamefont {Steil}, \citenamefont {Manmana}, \citenamefont {Sentef},
  \citenamefont {Reutzel},\ and\ \citenamefont {Mathias}}]{Merboldt2025}%
  \BibitemOpen
  \bibfield  {author} {\bibinfo {author} {\bibfnamefont {M.}~\bibnamefont
  {Merboldt}}, \bibinfo {author} {\bibfnamefont {M.}~\bibnamefont
  {Sch{\"u}ler}}, \bibinfo {author} {\bibfnamefont {D.}~\bibnamefont
  {Schmitt}}, \bibinfo {author} {\bibfnamefont {J.~P.}\ \bibnamefont {Bange}},
  \bibinfo {author} {\bibfnamefont {W.}~\bibnamefont {Bennecke}}, \bibinfo
  {author} {\bibfnamefont {K.}~\bibnamefont {Gadge}}, \bibinfo {author}
  {\bibfnamefont {K.}~\bibnamefont {Pierz}}, \bibinfo {author} {\bibfnamefont
  {H.~W.}\ \bibnamefont {Schumacher}}, \bibinfo {author} {\bibfnamefont
  {D.}~\bibnamefont {Momeni}}, \bibinfo {author} {\bibfnamefont
  {D.}~\bibnamefont {Steil}}, \bibinfo {author} {\bibfnamefont {S.~R.}\
  \bibnamefont {Manmana}}, \bibinfo {author} {\bibfnamefont {M.~A.}\
  \bibnamefont {Sentef}}, \bibinfo {author} {\bibfnamefont {M.}~\bibnamefont
  {Reutzel}},\ and\ \bibinfo {author} {\bibfnamefont {S.}~\bibnamefont
  {Mathias}},\ }\bibfield  {title} {\bibinfo {title} {Observation of floquet
  states in graphene},\ }\href {https://doi.org/10.1038/s41567-025-02889-7}
  {\bibfield  {journal} {\bibinfo  {journal} {Nat. Phys.}\ }\textbf {\bibinfo
  {volume} {21}},\ \bibinfo {pages} {1093} (\bibinfo {year}
  {2025})}\BibitemShut {NoStop}%
\bibitem [{\citenamefont {Wang}\ \emph {et~al.}(2026)\citenamefont {Wang},
  \citenamefont {Cai}, \citenamefont {Tang}, \citenamefont {Lu}, \citenamefont
  {Chen}, \citenamefont {Sheng}, \citenamefont {Feng}, \citenamefont {Zhong},
  \citenamefont {Zhang}, \citenamefont {Yu},\ and\ \citenamefont
  {Zhou}}]{Wang2026}%
  \BibitemOpen
  \bibfield  {author} {\bibinfo {author} {\bibfnamefont {F.}~\bibnamefont
  {Wang}}, \bibinfo {author} {\bibfnamefont {X.}~\bibnamefont {Cai}}, \bibinfo
  {author} {\bibfnamefont {X.}~\bibnamefont {Tang}}, \bibinfo {author}
  {\bibfnamefont {J.}~\bibnamefont {Lu}}, \bibinfo {author} {\bibfnamefont
  {W.}~\bibnamefont {Chen}}, \bibinfo {author} {\bibfnamefont {T.}~\bibnamefont
  {Sheng}}, \bibinfo {author} {\bibfnamefont {R.}~\bibnamefont {Feng}},
  \bibinfo {author} {\bibfnamefont {H.}~\bibnamefont {Zhong}}, \bibinfo
  {author} {\bibfnamefont {H.}~\bibnamefont {Zhang}}, \bibinfo {author}
  {\bibfnamefont {P.}~\bibnamefont {Yu}},\ and\ \bibinfo {author}
  {\bibfnamefont {S.}~\bibnamefont {Zhou}},\ }\bibfield  {title} {\bibinfo
  {title} {Observation of floquet-induced gap in graphene},\ }\bibfield
  {journal} {\bibinfo  {journal} {Nature Materials}\ }\href
  {https://doi.org/10.1038/s41563-026-02549-y} {10.1038/s41563-026-02549-y}
  (\bibinfo {year} {2026})\BibitemShut {NoStop}%
\bibitem [{\citenamefont {Zhang}\ \emph {et~al.}(2013)\citenamefont {Zhang},
  \citenamefont {Kane},\ and\ \citenamefont {Mele}}]{Zhang2013PRL}%
  \BibitemOpen
  \bibfield  {author} {\bibinfo {author} {\bibfnamefont {F.}~\bibnamefont
  {Zhang}}, \bibinfo {author} {\bibfnamefont {C.~L.}\ \bibnamefont {Kane}},\
  and\ \bibinfo {author} {\bibfnamefont {E.~J.}\ \bibnamefont {Mele}},\
  }\bibfield  {title} {\bibinfo {title} {Surface state magnetization and chiral
  edge states on topological insulators},\ }\href
  {https://doi.org/10.1103/PhysRevLett.110.046404} {\bibfield  {journal}
  {\bibinfo  {journal} {Phys. Rev. Lett.}\ }\textbf {\bibinfo {volume} {110}},\
  \bibinfo {pages} {046404} (\bibinfo {year} {2013})}\BibitemShut {NoStop}%
\bibitem [{\citenamefont {Benalcazar}\ \emph {et~al.}(2017)\citenamefont
  {Benalcazar}, \citenamefont {Bernevig},\ and\ \citenamefont
  {Hughes}}]{Benalcazar2017Science}%
  \BibitemOpen
  \bibfield  {author} {\bibinfo {author} {\bibfnamefont {W.~A.}\ \bibnamefont
  {Benalcazar}}, \bibinfo {author} {\bibfnamefont {B.~A.}\ \bibnamefont
  {Bernevig}},\ and\ \bibinfo {author} {\bibfnamefont {T.~L.}\ \bibnamefont
  {Hughes}},\ }\bibfield  {title} {\bibinfo {title} {Quantized electric
  multipole insulators},\ }\href {https://doi.org/10.1126/science.aah6442}
  {\bibfield  {journal} {\bibinfo  {journal} {Science}\ }\textbf {\bibinfo
  {volume} {357}},\ \bibinfo {pages} {61} (\bibinfo {year} {2017})}\BibitemShut
  {NoStop}%
\bibitem [{\citenamefont {Langbehn}\ \emph {et~al.}(2017)\citenamefont
  {Langbehn}, \citenamefont {Peng}, \citenamefont {Trifunovic}, \citenamefont
  {von Oppen},\ and\ \citenamefont {Brouwer}}]{Langbehn2017PRL}%
  \BibitemOpen
  \bibfield  {author} {\bibinfo {author} {\bibfnamefont {J.}~\bibnamefont
  {Langbehn}}, \bibinfo {author} {\bibfnamefont {Y.}~\bibnamefont {Peng}},
  \bibinfo {author} {\bibfnamefont {L.}~\bibnamefont {Trifunovic}}, \bibinfo
  {author} {\bibfnamefont {F.}~\bibnamefont {von Oppen}},\ and\ \bibinfo
  {author} {\bibfnamefont {P.~W.}\ \bibnamefont {Brouwer}},\ }\bibfield
  {title} {\bibinfo {title} {Reflection-symmetric second-order topological
  insulators and superconductors},\ }\href
  {https://doi.org/10.1103/PhysRevLett.119.246401} {\bibfield  {journal}
  {\bibinfo  {journal} {Phys. Rev. Lett.}\ }\textbf {\bibinfo {volume} {119}},\
  \bibinfo {pages} {246401} (\bibinfo {year} {2017})}\BibitemShut {NoStop}%
\bibitem [{\citenamefont {Song}\ \emph {et~al.}(2017)\citenamefont {Song},
  \citenamefont {Fang},\ and\ \citenamefont {Fang}}]{Song2017PRL}%
  \BibitemOpen
  \bibfield  {author} {\bibinfo {author} {\bibfnamefont {Z.}~\bibnamefont
  {Song}}, \bibinfo {author} {\bibfnamefont {Z.}~\bibnamefont {Fang}},\ and\
  \bibinfo {author} {\bibfnamefont {C.}~\bibnamefont {Fang}},\ }\bibfield
  {title} {\bibinfo {title} {$(d\ensuremath{-}2)$-dimensional edge states of
  rotation symmetry protected topological states},\ }\href
  {https://doi.org/10.1103/PhysRevLett.119.246402} {\bibfield  {journal}
  {\bibinfo  {journal} {Phys. Rev. Lett.}\ }\textbf {\bibinfo {volume} {119}},\
  \bibinfo {pages} {246402} (\bibinfo {year} {2017})}\BibitemShut {NoStop}%
\bibitem [{\citenamefont {Schindler}\ \emph {et~al.}(2018)\citenamefont
  {Schindler}, \citenamefont {Cook}, \citenamefont {Vergniory}, \citenamefont
  {Wang}, \citenamefont {Parkin}, \citenamefont {Bernevig},\ and\ \citenamefont
  {Neupert}}]{Schindler2018SA}%
  \BibitemOpen
  \bibfield  {author} {\bibinfo {author} {\bibfnamefont {F.}~\bibnamefont
  {Schindler}}, \bibinfo {author} {\bibfnamefont {A.~M.}\ \bibnamefont {Cook}},
  \bibinfo {author} {\bibfnamefont {M.~G.}\ \bibnamefont {Vergniory}}, \bibinfo
  {author} {\bibfnamefont {Z.}~\bibnamefont {Wang}}, \bibinfo {author}
  {\bibfnamefont {S.~S.~P.}\ \bibnamefont {Parkin}}, \bibinfo {author}
  {\bibfnamefont {B.~A.}\ \bibnamefont {Bernevig}},\ and\ \bibinfo {author}
  {\bibfnamefont {T.}~\bibnamefont {Neupert}},\ }\bibfield  {title} {\bibinfo
  {title} {Higher-order topological insulators},\ }\href
  {https://doi.org/10.1126/sciadv.aat0346} {\bibfield  {journal} {\bibinfo
  {journal} {Sci. Adv.}\ }\textbf {\bibinfo {volume} {4}},\ \bibinfo {pages}
  {eaat0346} (\bibinfo {year} {2018})}\BibitemShut {NoStop}%
\bibitem [{\citenamefont {Chen}\ \emph {et~al.}(2020)\citenamefont {Chen},
  \citenamefont {Chen}, \citenamefont {Gao}, \citenamefont {Zhou},\ and\
  \citenamefont {Xu}}]{PhysRevLett.124.036803}%
  \BibitemOpen
  \bibfield  {author} {\bibinfo {author} {\bibfnamefont {R.}~\bibnamefont
  {Chen}}, \bibinfo {author} {\bibfnamefont {C.-Z.}\ \bibnamefont {Chen}},
  \bibinfo {author} {\bibfnamefont {J.-H.}\ \bibnamefont {Gao}}, \bibinfo
  {author} {\bibfnamefont {B.}~\bibnamefont {Zhou}},\ and\ \bibinfo {author}
  {\bibfnamefont {D.-H.}\ \bibnamefont {Xu}},\ }\bibfield  {title} {\bibinfo
  {title} {Higher-order topological insulators in quasicrystals},\ }\href
  {https://doi.org/10.1103/PhysRevLett.124.036803} {\bibfield  {journal}
  {\bibinfo  {journal} {Phys. Rev. Lett.}\ }\textbf {\bibinfo {volume} {124}},\
  \bibinfo {pages} {036803} (\bibinfo {year} {2020})}\BibitemShut {NoStop}%
\bibitem [{\citenamefont {Xie}\ \emph {et~al.}(2021)\citenamefont {Xie},
  \citenamefont {Wang}, \citenamefont {Zhang}, \citenamefont {Zhan},
  \citenamefont {Jiang}, \citenamefont {Lu},\ and\ \citenamefont
  {Chen}}]{xie2021higher}%
  \BibitemOpen
  \bibfield  {author} {\bibinfo {author} {\bibfnamefont {B.}~\bibnamefont
  {Xie}}, \bibinfo {author} {\bibfnamefont {H.-X.}\ \bibnamefont {Wang}},
  \bibinfo {author} {\bibfnamefont {X.}~\bibnamefont {Zhang}}, \bibinfo
  {author} {\bibfnamefont {P.}~\bibnamefont {Zhan}}, \bibinfo {author}
  {\bibfnamefont {J.-H.}\ \bibnamefont {Jiang}}, \bibinfo {author}
  {\bibfnamefont {M.}~\bibnamefont {Lu}},\ and\ \bibinfo {author}
  {\bibfnamefont {Y.}~\bibnamefont {Chen}},\ }\bibfield  {title} {\bibinfo
  {title} {Higher-order band topology},\ }\href
  {https://doi.org/10.1038/s42254-021-00323-4} {\bibfield  {journal} {\bibinfo
  {journal} {Nat. Rev. Phys.}\ }\textbf {\bibinfo {volume} {3}},\ \bibinfo
  {pages} {520} (\bibinfo {year} {2021})}\BibitemShut {NoStop}%
\bibitem [{\citenamefont {Si}\ \emph {et~al.}(2016)\citenamefont {Si},
  \citenamefont {Sun},\ and\ \citenamefont {Liu}}]{C5NR07755A}%
  \BibitemOpen
  \bibfield  {author} {\bibinfo {author} {\bibfnamefont {C.}~\bibnamefont
  {Si}}, \bibinfo {author} {\bibfnamefont {Z.}~\bibnamefont {Sun}},\ and\
  \bibinfo {author} {\bibfnamefont {F.}~\bibnamefont {Liu}},\ }\bibfield
  {title} {\bibinfo {title} {Strain engineering of graphene: a review},\ }\href
  {https://doi.org/10.1039/C5NR07755A} {\bibfield  {journal} {\bibinfo
  {journal} {Nanoscale}\ }\textbf {\bibinfo {volume} {8}},\ \bibinfo {pages}
  {3207} (\bibinfo {year} {2016})}\BibitemShut {NoStop}%
\bibitem [{\citenamefont {Choi}\ \emph {et~al.}(2010)\citenamefont {Choi},
  \citenamefont {Jhi},\ and\ \citenamefont {Son}}]{PhysRevB.81.081407}%
  \BibitemOpen
  \bibfield  {author} {\bibinfo {author} {\bibfnamefont {S.-M.}\ \bibnamefont
  {Choi}}, \bibinfo {author} {\bibfnamefont {S.-H.}\ \bibnamefont {Jhi}},\ and\
  \bibinfo {author} {\bibfnamefont {Y.-W.}\ \bibnamefont {Son}},\ }\bibfield
  {title} {\bibinfo {title} {Effects of strain on electronic properties of
  graphene},\ }\href {https://doi.org/10.1103/PhysRevB.81.081407} {\bibfield
  {journal} {\bibinfo  {journal} {Phys. Rev. B}\ }\textbf {\bibinfo {volume}
  {81}},\ \bibinfo {pages} {081407} (\bibinfo {year} {2010})}\BibitemShut
  {NoStop}%
\bibitem [{\citenamefont {Naumis}\ \emph {et~al.}(2017)\citenamefont {Naumis},
  \citenamefont {Barraza-Lopez}, \citenamefont {Oliva-Leyva},\ and\
  \citenamefont {Terrones}}]{Naumis_2017}%
  \BibitemOpen
  \bibfield  {author} {\bibinfo {author} {\bibfnamefont {G.~G.}\ \bibnamefont
  {Naumis}}, \bibinfo {author} {\bibfnamefont {S.}~\bibnamefont
  {Barraza-Lopez}}, \bibinfo {author} {\bibfnamefont {M.}~\bibnamefont
  {Oliva-Leyva}},\ and\ \bibinfo {author} {\bibfnamefont {H.}~\bibnamefont
  {Terrones}},\ }\bibfield  {title} {\bibinfo {title} {Electronic and optical
  properties of strained graphene and other strained 2d materials: a review},\
  }\href {https://doi.org/10.1088/1361-6633/aa74ef} {\bibfield  {journal}
  {\bibinfo  {journal} {Reports on Progress in Physics}\ }\textbf {\bibinfo
  {volume} {80}},\ \bibinfo {pages} {096501} (\bibinfo {year}
  {2017})}\BibitemShut {NoStop}%
\bibitem [{\citenamefont {Pereira}\ \emph {et~al.}(2011)\citenamefont
  {Pereira}, \citenamefont {Ribeiro}, \citenamefont {Peres},\ and\
  \citenamefont {Castro~Neto}}]{Pereira_2010}%
  \BibitemOpen
  \bibfield  {author} {\bibinfo {author} {\bibfnamefont {V.~M.}\ \bibnamefont
  {Pereira}}, \bibinfo {author} {\bibfnamefont {R.~M.}\ \bibnamefont
  {Ribeiro}}, \bibinfo {author} {\bibfnamefont {N.~M.~R.}\ \bibnamefont
  {Peres}},\ and\ \bibinfo {author} {\bibfnamefont {A.~H.}\ \bibnamefont
  {Castro~Neto}},\ }\bibfield  {title} {\bibinfo {title} {Optical properties of
  strained graphene},\ }\href {https://doi.org/10.1209/0295-5075/92/67001}
  {\bibfield  {journal} {\bibinfo  {journal} {Europhysics Letters}\ }\textbf
  {\bibinfo {volume} {92}},\ \bibinfo {pages} {67001} (\bibinfo {year}
  {2011})}\BibitemShut {NoStop}%
\bibitem [{\citenamefont {Gui}\ \emph {et~al.}(2008)\citenamefont {Gui},
  \citenamefont {Li},\ and\ \citenamefont {Zhong}}]{PhysRevB.78.075435}%
  \BibitemOpen
  \bibfield  {author} {\bibinfo {author} {\bibfnamefont {G.}~\bibnamefont
  {Gui}}, \bibinfo {author} {\bibfnamefont {J.}~\bibnamefont {Li}},\ and\
  \bibinfo {author} {\bibfnamefont {J.}~\bibnamefont {Zhong}},\ }\bibfield
  {title} {\bibinfo {title} {Band structure engineering of graphene by strain:
  First-principles calculations},\ }\href
  {https://doi.org/10.1103/PhysRevB.78.075435} {\bibfield  {journal} {\bibinfo
  {journal} {Phys. Rev. B}\ }\textbf {\bibinfo {volume} {78}},\ \bibinfo
  {pages} {075435} (\bibinfo {year} {2008})}\BibitemShut {NoStop}%
\bibitem [{\citenamefont {Lee}\ \emph {et~al.}(2008)\citenamefont {Lee},
  \citenamefont {Wei}, \citenamefont {Kysar},\ and\ \citenamefont
  {Hone}}]{Lee2008Sci}%
  \BibitemOpen
  \bibfield  {author} {\bibinfo {author} {\bibfnamefont {C.}~\bibnamefont
  {Lee}}, \bibinfo {author} {\bibfnamefont {X.}~\bibnamefont {Wei}}, \bibinfo
  {author} {\bibfnamefont {J.~W.}\ \bibnamefont {Kysar}},\ and\ \bibinfo
  {author} {\bibfnamefont {J.}~\bibnamefont {Hone}},\ }\bibfield  {title}
  {\bibinfo {title} {Measurement of the elastic properties and intrinsic
  strength of monolayer graphene},\ }\href
  {https://doi.org/10.1126/science.1157996} {\bibfield  {journal} {\bibinfo
  {journal} {Science}\ }\textbf {\bibinfo {volume} {321}},\ \bibinfo {pages}
  {385} (\bibinfo {year} {2008})}\BibitemShut {NoStop}%
\bibitem [{\citenamefont {Wei}\ \emph {et~al.}(2009)\citenamefont {Wei},
  \citenamefont {Fragneaud}, \citenamefont {Marianetti},\ and\ \citenamefont
  {Kysar}}]{Wei2009PRB}%
  \BibitemOpen
  \bibfield  {author} {\bibinfo {author} {\bibfnamefont {X.}~\bibnamefont
  {Wei}}, \bibinfo {author} {\bibfnamefont {B.}~\bibnamefont {Fragneaud}},
  \bibinfo {author} {\bibfnamefont {C.~A.}\ \bibnamefont {Marianetti}},\ and\
  \bibinfo {author} {\bibfnamefont {J.~W.}\ \bibnamefont {Kysar}},\ }\bibfield
  {title} {\bibinfo {title} {Nonlinear elastic behavior of graphene: Ab initio
  calculations to continuum description},\ }\href
  {https://doi.org/10.1103/PhysRevB.80.205407} {\bibfield  {journal} {\bibinfo
  {journal} {Phys. Rev. B}\ }\textbf {\bibinfo {volume} {80}},\ \bibinfo
  {pages} {205407} (\bibinfo {year} {2009})}\BibitemShut {NoStop}%
\bibitem [{\citenamefont {Dietl}\ \emph {et~al.}(2008)\citenamefont {Dietl},
  \citenamefont {Pi\'echon},\ and\ \citenamefont {Montambaux}}]{SemiDirac1}%
  \BibitemOpen
  \bibfield  {author} {\bibinfo {author} {\bibfnamefont {P.}~\bibnamefont
  {Dietl}}, \bibinfo {author} {\bibfnamefont {F.}~\bibnamefont {Pi\'echon}},\
  and\ \bibinfo {author} {\bibfnamefont {G.}~\bibnamefont {Montambaux}},\
  }\bibfield  {title} {\bibinfo {title} {New magnetic field dependence of
  landau levels in a graphenelike structure},\ }\href
  {https://doi.org/10.1103/PhysRevLett.100.236405} {\bibfield  {journal}
  {\bibinfo  {journal} {Phys. Rev. Lett.}\ }\textbf {\bibinfo {volume} {100}},\
  \bibinfo {pages} {236405} (\bibinfo {year} {2008})}\BibitemShut {NoStop}%
\bibitem [{\citenamefont {Banerjee}\ \emph {et~al.}(2009)\citenamefont
  {Banerjee}, \citenamefont {Singh}, \citenamefont {Pardo},\ and\ \citenamefont
  {Pickett}}]{SemiDirac2}%
  \BibitemOpen
  \bibfield  {author} {\bibinfo {author} {\bibfnamefont {S.}~\bibnamefont
  {Banerjee}}, \bibinfo {author} {\bibfnamefont {R.~R.~P.}\ \bibnamefont
  {Singh}}, \bibinfo {author} {\bibfnamefont {V.}~\bibnamefont {Pardo}},\ and\
  \bibinfo {author} {\bibfnamefont {W.~E.}\ \bibnamefont {Pickett}},\
  }\bibfield  {title} {\bibinfo {title} {Tight-binding modeling and low-energy
  behavior of the semi-dirac point},\ }\href
  {https://doi.org/10.1103/PhysRevLett.103.016402} {\bibfield  {journal}
  {\bibinfo  {journal} {Phys. Rev. Lett.}\ }\textbf {\bibinfo {volume} {103}},\
  \bibinfo {pages} {016402} (\bibinfo {year} {2009})}\BibitemShut {NoStop}%
\bibitem [{\citenamefont {Shirley}(1965)}]{PhysRev.138.B979}%
  \BibitemOpen
  \bibfield  {author} {\bibinfo {author} {\bibfnamefont {J.~H.}\ \bibnamefont
  {Shirley}},\ }\bibfield  {title} {\bibinfo {title} {Solution of the
  {S}chr\"odinger equation with a hamiltonian periodic in time},\ }\href
  {https://doi.org/10.1103/PhysRev.138.B979} {\bibfield  {journal} {\bibinfo
  {journal} {Phys. Rev.}\ }\textbf {\bibinfo {volume} {138}},\ \bibinfo {pages}
  {B979} (\bibinfo {year} {1965})}\BibitemShut {NoStop}%
\bibitem [{\citenamefont {Sambe}(1973)}]{PhysRevA.7.2203}%
  \BibitemOpen
  \bibfield  {author} {\bibinfo {author} {\bibfnamefont {H.}~\bibnamefont
  {Sambe}},\ }\bibfield  {title} {\bibinfo {title} {Steady states and
  quasienergies of a quantum-mechanical system in an oscillating field},\
  }\href {https://doi.org/10.1103/PhysRevA.7.2203} {\bibfield  {journal}
  {\bibinfo  {journal} {Phys. Rev. A}\ }\textbf {\bibinfo {volume} {7}},\
  \bibinfo {pages} {2203} (\bibinfo {year} {1973})}\BibitemShut {NoStop}%
\bibitem [{\citenamefont {Bukov}\ \emph {et~al.}(2015)\citenamefont {Bukov},
  \citenamefont {D'Alessio},\ and\ \citenamefont
  {Polkovnikov}}]{bukov2015universal}%
  \BibitemOpen
  \bibfield  {author} {\bibinfo {author} {\bibfnamefont {M.}~\bibnamefont
  {Bukov}}, \bibinfo {author} {\bibfnamefont {L.}~\bibnamefont {D'Alessio}},\
  and\ \bibinfo {author} {\bibfnamefont {A.}~\bibnamefont {Polkovnikov}},\
  }\bibfield  {title} {\bibinfo {title} {Universal high-frequency behavior of
  periodically driven systems: from dynamical stabilization to {F}loquet
  engineering},\ }\href {https://doi.org/10.1080/00018732.2015.1055918}
  {\bibfield  {journal} {\bibinfo  {journal} {Adv. Phys.}\ }\textbf {\bibinfo
  {volume} {64}},\ \bibinfo {pages} {139} (\bibinfo {year} {2015})}\BibitemShut
  {NoStop}%
\bibitem [{\citenamefont {Eckardt}\ and\ \citenamefont
  {Anisimovas}(2015)}]{Eckardt_2015}%
  \BibitemOpen
  \bibfield  {author} {\bibinfo {author} {\bibfnamefont {A.}~\bibnamefont
  {Eckardt}}\ and\ \bibinfo {author} {\bibfnamefont {E.}~\bibnamefont
  {Anisimovas}},\ }\bibfield  {title} {\bibinfo {title} {High-frequency
  approximation for periodically driven quantum systems from a {F}loquet-space
  perspective},\ }\href {https://doi.org/10.1088/1367-2630/17/9/093039}
  {\bibfield  {journal} {\bibinfo  {journal} {New J. Phys.}\ }\textbf {\bibinfo
  {volume} {17}},\ \bibinfo {pages} {093039} (\bibinfo {year}
  {2015})}\BibitemShut {NoStop}%
\bibitem [{\citenamefont {Ezawa}(2018)}]{Ezawa2018PRL}%
  \BibitemOpen
  \bibfield  {author} {\bibinfo {author} {\bibfnamefont {M.}~\bibnamefont
  {Ezawa}},\ }\bibfield  {title} {\bibinfo {title} {Higher-order topological
  insulators and semimetals on the breathing {Kagome} and pyrochlore
  lattices},\ }\href {https://doi.org/10.1103/PhysRevLett.120.026801}
  {\bibfield  {journal} {\bibinfo  {journal} {Phys. Rev. Lett.}\ }\textbf
  {\bibinfo {volume} {120}},\ \bibinfo {pages} {026801} (\bibinfo {year}
  {2018})}\BibitemShut {NoStop}%
\bibitem [{\citenamefont {Hohenberg}\ and\ \citenamefont
  {Kohn}(1964)}]{Hohenberg1964}%
  \BibitemOpen
  \bibfield  {author} {\bibinfo {author} {\bibfnamefont {P.}~\bibnamefont
  {Hohenberg}}\ and\ \bibinfo {author} {\bibfnamefont {W.}~\bibnamefont
  {Kohn}},\ }\bibfield  {title} {\bibinfo {title} {Inhomogeneous electron
  gas},\ }\href {https://doi.org/10.1103/PhysRev.136.B864} {\bibfield
  {journal} {\bibinfo  {journal} {Phys. Rev.}\ }\textbf {\bibinfo {volume}
  {136}},\ \bibinfo {pages} {B864} (\bibinfo {year} {1964})}\BibitemShut
  {NoStop}%
\bibitem [{\citenamefont {Kohn}\ and\ \citenamefont {Sham}(1965)}]{Kohn1965}%
  \BibitemOpen
  \bibfield  {author} {\bibinfo {author} {\bibfnamefont {W.}~\bibnamefont
  {Kohn}}\ and\ \bibinfo {author} {\bibfnamefont {L.~J.}\ \bibnamefont
  {Sham}},\ }\bibfield  {title} {\bibinfo {title} {Self-consistent equations
  including exchange and correlation effects},\ }\href
  {https://doi.org/10.1103/PhysRev.140.A1133} {\bibfield  {journal} {\bibinfo
  {journal} {Phys. Rev.}\ }\textbf {\bibinfo {volume} {140}},\ \bibinfo {pages}
  {A1133} (\bibinfo {year} {1965})}\BibitemShut {NoStop}%
\bibitem [{\citenamefont {Kresse}\ and\ \citenamefont
  {Furthm\"uller}(1996)}]{Kresse1996}%
  \BibitemOpen
  \bibfield  {author} {\bibinfo {author} {\bibfnamefont {G.}~\bibnamefont
  {Kresse}}\ and\ \bibinfo {author} {\bibfnamefont {J.}~\bibnamefont
  {Furthm\"uller}},\ }\bibfield  {title} {\bibinfo {title} {Efficient iterative
  schemes for ab initio total-energy calculations using a plane-wave basis
  set},\ }\href {https://doi.org/10.1103/PhysRevB.54.11169} {\bibfield
  {journal} {\bibinfo  {journal} {Phys. Rev. B}\ }\textbf {\bibinfo {volume}
  {54}},\ \bibinfo {pages} {11169} (\bibinfo {year} {1996})}\BibitemShut
  {NoStop}%
\bibitem [{\citenamefont {Perdew}\ \emph {et~al.}(1996)\citenamefont {Perdew},
  \citenamefont {Burke},\ and\ \citenamefont {Ernzerhof}}]{Perdew1996}%
  \BibitemOpen
  \bibfield  {author} {\bibinfo {author} {\bibfnamefont {J.~P.}\ \bibnamefont
  {Perdew}}, \bibinfo {author} {\bibfnamefont {K.}~\bibnamefont {Burke}},\ and\
  \bibinfo {author} {\bibfnamefont {M.}~\bibnamefont {Ernzerhof}},\ }\bibfield
  {title} {\bibinfo {title} {Generalized gradient approximation made simple},\
  }\href {https://doi.org/10.1103/PhysRevLett.77.3865} {\bibfield  {journal}
  {\bibinfo  {journal} {Phys. Rev. Lett.}\ }\textbf {\bibinfo {volume} {77}},\
  \bibinfo {pages} {3865} (\bibinfo {year} {1996})}\BibitemShut {NoStop}%
\bibitem [{\citenamefont {Wu}\ \emph {et~al.}(2018)\citenamefont {Wu},
  \citenamefont {Zhang}, \citenamefont {Song}, \citenamefont {Troyer},\ and\
  \citenamefont {Soluyanov}}]{WU2018405}%
  \BibitemOpen
  \bibfield  {author} {\bibinfo {author} {\bibfnamefont {Q.}~\bibnamefont
  {Wu}}, \bibinfo {author} {\bibfnamefont {S.}~\bibnamefont {Zhang}}, \bibinfo
  {author} {\bibfnamefont {H.-F.}\ \bibnamefont {Song}}, \bibinfo {author}
  {\bibfnamefont {M.}~\bibnamefont {Troyer}},\ and\ \bibinfo {author}
  {\bibfnamefont {A.~A.}\ \bibnamefont {Soluyanov}},\ }\bibfield  {title}
  {\bibinfo {title} {Wanniertools: An open-source software package for novel
  topological materials},\ }\href
  {https://doi.org/https://doi.org/10.1016/j.cpc.2017.09.033} {\bibfield
  {journal} {\bibinfo  {journal} {Comput. Phys. Commun.}\ }\textbf {\bibinfo
  {volume} {224}},\ \bibinfo {pages} {405} (\bibinfo {year}
  {2018})}\BibitemShut {NoStop}%
\end{thebibliography}%

\end{document}